\documentclass[a4paper,10pt]{article}

\usepackage{graphicx}
\usepackage{cite}
\usepackage{a4wide}
\usepackage{amsthm}
\usepackage{amsmath}
\usepackage{amssymb}
\usepackage{caption}
\usepackage{subfig,xcolor}
\usepackage[export]{adjustbox}
\usepackage{float}

\newcommand{\rot}[1]{\mathcal{R}_{\{#1\}}}
\newcommand{\boost}[1]{\mathcal{B}_{\{#1\}}}
\newcommand{\cstring}[1]{\mathcal{S}_{\{#1\}}}

\newcommand{\cV}{V}

\def \im {{i}}
\def \ii {{\rm i}}

\def \dd {{\rm d}}
\def \U {{\cal U}}
\def \V {{\cal V}}
\def \Up {{U_{+}}}

\title{An interpretation of spacetimes with expanding impulsive gravitational waves generated by snapped cosmic strings}
\author{
D.~Kofro{\v n}$^1$\thanks{{\tt d.kofron@gmail.com}},
M.~Karamazov$^1$\thanks{{\tt michal.karamazov@gmail.com}},
and R.~\v{S}varc$^1$\thanks{{\tt robert.svarc@mff.cuni.cz}} \\ \\
$^1$ Institute of Theoretical Physics,\\
Charles University, Faculty of Mathematics and Physics,\\
V Hole\v{s}ovi\v{c}k\'ach 2, 180\,00 Prague 8, Czech Republic.\\ \\
}
\date{}

\begin{document}

\maketitle

\begin{abstract}
The geometric properties of spacetimes representing expanding impulsive gravitational waves, propagating on a flat background and generated by snapped cosmic strings, are studied. The construction of the line element is reviewed, and suitable forms of the string-generating complex mapping are derived for various configurations such as previously studied examples of a pair of snapping cosmic strings. Moreover, these mappings are related to the topology of the flat half-space in front of the wave. Their understanding seems to be crucial for further analysis of the global geometry, the relation between half-spaces on both sides of the impulse, and the physical interpretation of, in principle, observable effects. The spacetime structure is connected with the motion of free test particles crossing the impulse, where the recent results allow us to discuss their displacement and induced velocities that are caused by the interaction with the expanding gravitational impulsive wave.
\end{abstract}

\section{Introduction}

Interestingly, the prediction of gravitational waves (GW) became an immediate consequence of Einstein's geometric description of gravity in terms of curved spacetime. The analysis of weak field regime \cite{Einstein16,Einstein18} as well as pioneering studies of invariants within full theory \cite{Piarni:1957, Bondi:1957} pointed out their inherent properties and gave quantitative estimates about the magnitude of typical amplitudes. Simultaneously, it became clear that their observation is going to be very challenging or even impossible. In 1974 Hulse and Taylor discovered the binary pulsar PSR 1913+16 \cite{HulseTaylor75}. The systematic measurements of its orbital period shifts exactly corresponded to the gravitational energy loss radiated away in the form of GWs. This was in perfect agreement with the general relativity (GR) prediction and endorsed the hunt for their direct detection. The unrelenting effort was rewarded in 2015 when the LIGO facility detected a signal produced by merging pair of black holes more than one billion light-years away \cite{Abbott16}. Until now, dozens of other events have been detected with the prominent case of merging neutron stars \cite{Abbott17}, where the GW signal was supplemented by observations in the whole range of the electromagnetic spectrum.

These ultimate experimental achievements are based on sophisticated numerical simulations. However, to better understand the GW properties, or to verify various models and numerical schemes, even the exact analytical radiative spacetimes are necessary. The most important wave-like solutions to Einstein's general relativity belong to the expanding Robinson--Trautman and non-expanding Kundt classes of exact spacetimes \cite{RobTra60, RobTra62, Kundt:1961, Kundt:1962}, see \cite{Stephanietal:book, GriffithsPodolsky:book} for the comprehensive review. Interestingly, both these classes allow for the \emph{impulsive} profiles of the wave, where the propagating curvature is located only on the singular null hypersurface. Technically, it corresponds to the presence of distributional terms in the spacetime geometric quantities. Therefore, the impulsive solutions are interpreted as extremely short, but simultaneously very intense, bursts of gravitational radiation. Surprisingly, these geometries are also of a purely mathematical interest since careful manipulation with inherently non-linear distributional terms is required that necessarily goes beyond the classical GR, see e.g. \cite{Steinbauer:book, SteinbauerVickers:2006}. 

Within this paper, we are interested in particular aspects of expanding Robinson--Trautman impulses propagating on a flat Minkowski background. It is worth mentioning that the elegant geometric construction of these models goes back to the seventies when Roger Penrose proposed the `cut and paste' approach \cite{Penrose72}. Two decades later Penrose and Nutku found the continuous line element for these solutions \cite{NutkuPenrose92}. Subsequently, various extensions of such a construction were presented, e.g, including the cosmological constant or other additional parameters \cite{Hogan92, PodolskyGriffiths99, Hogan95, AlievNutku2001}. The historical content, summary of other construction methods (e.g., limits of expanding sandwich waves or infinite acceleration limit of the C-metric), and detailed list of references can be found, e.g., in \cite{GriffithsPodolsky:book, Podolsky:chapter, PodolskySamannSteinbauerSvarc2016}. In the original work \cite{NutkuPenrose92}, the snapped cosmic string serving as a wave source was described and the possibility of impulses generated by a pair of colliding and snapping strings was outlined. This scenario was elaborated by Podolsk\'y and Griffiths \cite{PodolskyGriffiths00}. However, the complete understanding of the spacetime topology remained an open problem. Simultaneously, the topology plays a crucial role in the analysis of a geodesic motion affected by the induced impulsive wave, and vice versa, the geodesic motion reflects geometry of the impulse. Therefore, our present contribution aims to fill up this gap and extend discussion of the wave sources. In section \ref{Sec:ImpConstruction}, the description of expanding impulsive waves is summarized. Subsequently, the construction of string-like geometries is reviewed in section~\ref{Sec:ImpGeometry} and the description of geodesics is presented in section~\ref{Sec:Geodesics}. In the last two sections~\ref{Sec:OneString} and~\ref{Sec:TwoStrings}, we discuss and interpret specific properties of one and two-string geometries, respectively, analysing the interaction of corresponding gravitational impulses with free test particles.    

\section{Expanding gravitational impulses on a flat background\label{Sec:ImpConstruction}}

It is natural to begin with a description of the Penrose geometric `cut and paste' construction in the simplest situation of an expanding impulse propagating on a flat Minkowski background
\begin{equation}
\dd s^2=-\dd t^2+\dd x^2+\dd y^2+\dd z^2 \,, \label{CartMink}
\end{equation}
which can be simply rewritten using the double null coordinates,
\begin{equation}
t=\frac{1}{\sqrt{2}}(\V+\U) \,, \qquad z=\frac{1}{\sqrt{2}}(\V-\U) \,, \qquad x=\frac{1}{\sqrt{2}}(\eta+\bar{\eta}) \,, \qquad y=\frac{1}{\im\sqrt{2}}(\eta-\bar{\eta}) \,, \label{MinkToNull}
\end{equation}
to the form
\begin{equation}
\dd s^2=-2\dd\U\dd\V+2\dd\eta\dd\bar{\eta} \,. \label{ExpDoubleNullMink}
\end{equation}
Applying further transformation to the Minkowski metric (\ref{ExpDoubleNullMink}),
\begin{equation}
\V = \frac{V}{p}-\epsilon U \,, \qquad \U = \frac{|Z|^2}{p}\,V-U\,, \qquad \eta\,= \frac{Z}{p}\,V  \label{NullConeTrans}
\end{equation}
with
\begin{equation}
p=1+\epsilon Z\bar Z\,, \qquad \hbox{where} \qquad \epsilon=-1,0,+1 \,,
\end{equation}
we get the line element
\begin{equation}
\dd s^2 = 2\,\frac{V^2}{p^2}\,\dd Z\dd\bar Z +2\,\dd U\dd V -2\epsilon\,\dd U^2 \,, \label{NullConeMetric}
\end{equation}
which explicitly describes the foliation of the flat spacetime by null cones that are labelled by  constant values of the coordinate $U$. The parameter $\epsilon$ encodes the Gaussian curvature of spatial two-surfaces ${U=\mbox{const}}$ and ${V=\mbox{const}}$, see \cite{GriffithsPodolsky:book} for the detailed geometric picture.

However, another more involved transformation of (\ref{ExpDoubleNullMink}) can be performed (simultaneously leading to the explicit null-cone foliation), namely
\begin{align}
\V = AV-DU\,,   \qquad \U = BV-EU\,,   \qquad \eta\,= CV-FU\,, \label{ContTrans+}
\end{align}
where
\begin{align}
A=& \frac{1}{p|h'|}\,, & D=& \frac{1}{|h'|}\left\{\frac{p}{4} \left|\frac{h''}{h'}\right|^2+\epsilon\left[1+\frac{Z}{2}\frac{h''}{h'}+\frac{\bar Z}{2}\frac{\bar h''}{\bar h'}\right]\right\}, \nonumber\\
B=& \frac{|h|^2}{p|h'|}\,, & E=& \frac{|h|^2}{|h'|}\bigg\{ \frac{p}{4}\left|\frac{h''}{h'}-2\frac{h'}{h}\right|^2+\epsilon\left[ 1+\frac{Z}{2}\left(\frac{h''}{h'}-2\frac{h'}{h}\right)+\frac{\bar Z}{2}\left(\frac{\bar h''}{\bar h'}-2\frac{\bar h'}{\bar h}\right)\right]\bigg\}, \label{CoeffContTrans+} \\
C=& \frac{h}{ p|h'|}\,, & F=& \frac{h}{|h'|}\bigg\{\frac{p}{4}\left(\frac{h''}{h'}-2\frac{h'}{h}\right)\frac{\bar h''}{\bar h'}+\epsilon\left[1+\frac{Z}{2}\left(\frac{h''}{h'}-2\frac{h'}{h}\right)+\frac{\bar Z}{2}\frac{\bar h''}{\bar h'}\right]\bigg\}, \nonumber
\end{align}	
with ${h=h(Z)}$ representing an arbitrary complex holomorphic function (apart from its singular points) and a prime denoting its derivative. The resulting line element becomes
\begin{equation}
\dd s^2 = 2\left|\frac{V}{p}\,\dd Z+ U \,p\bar H\,\dd\bar Z \right|^2 +2\,\dd U\dd V -2\epsilon\,\dd U^2 \,, \label{ContMetric+}
\end{equation}
where $H$ is the Schwarzian derivative of the function $h$ defined as
\begin{equation}
H(h(Z))\equiv\frac{1}{2} \left[\frac{h'''}{h'}-\frac{3}{2}\left(\frac{h''}{h'}\right)^2\right] . \label{SchwarzDer}
\end{equation}
Although the metric (\ref{ContMetric+}) still represents the flat space, the non-triviality of function $H$ leads to the topological defects primarily induced by the choice of its generator $h(Z)$.

\begin{figure}[htb]
\begin{center}
\includegraphics[scale=0.95]{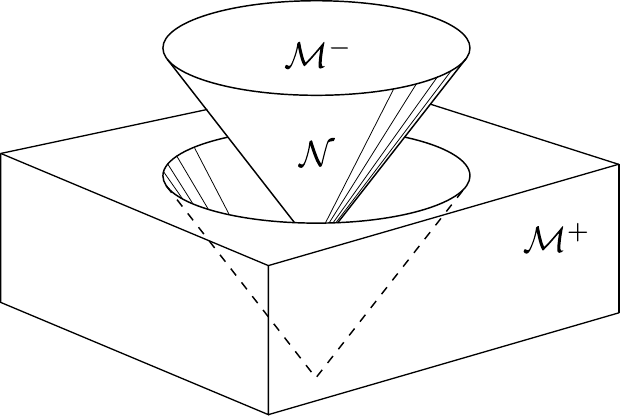}
\caption{\small Two parts ${\cal M}^-$ and ${\cal M}^+$ of the flat Minkowski background are re-attached along a future-oriented null cone~${\cal N}$ with a suitable warp given by the junction Penrose conditions (\ref{JuncCond}). Such construction results in the presence of expanding spherical gravitational impulse located on ${\cal N}$ whose specific nature is encoded in the (holomorphic) function ${h(Z)}$ identifying points of ${\cal M}^-$ and ${\cal M}^+$ across~${\cal N}$.}
\label{CutAndPAste}
\end{center}    
\end{figure}

Finally, to construct the expanding impulsive gravitational wave on a flat background one has to cut the Minkowski spacetime along the null cone and then re-attached the two half-spaces with an appropriate warp, see figure~\ref{CutAndPAste}. Using the coordinates of (\ref{NullConeMetric}) and taking the null cone ${U=0}$, corresponding to the expanding sphere ${t^2=x^2+y^2+z^2}$, the half-spaces ${{\cal M}^-}$ (with ${U\leq0}$) and ${{\cal M}^+}$ (with ${U\geq0}$) have to be identified across the null hypersurface ${\cal N}$ as
\begin{equation}
\left[Z,\,\bar Z,\,V,\,U=0_-\right]_{_{{\cal M}^-}} \equiv \left[h(Z),\,\bar h({\bar Z}),\,\frac{1+\epsilon\, h\bar h}{1+\epsilon Z\bar Z} \frac{V}{|h'|},\,U=0_+\right]_{_{{\cal M}^+}}\,.  \label{JuncCond}
\end{equation}

The Penrose junction conditions directly correspond to the evaluation of transformations (\ref{NullConeTrans}) and (\ref{ContTrans+}), respectively, on the impulse ${U=0}$, i.e., ${\U\V-\eta\bar{\eta}=0}$. The global \emph{continuous} line element can be then written as a combination of (\ref{NullConeMetric}) and (\ref{ContMetric+}), namely
\begin{equation}
\dd s^2 = 2\left|\frac{V}{p}\,\dd Z+ U\Theta(U) \,p\bar H\,\dd\bar Z \right|^2 +2\,\dd U\dd V -2\epsilon\,\dd U^2 \,, \label{ContMetric}
\end{equation}
where the product of $U$ and Heaviside step $\Theta(U)$ represents the continuous kink function which is typically denoted  as ${\Up\equiv U\Theta(U)}$. The metric (\ref{ContMetric}) then solves the vacuum Einstein field equations everywhere except at the singular impulse origin (${U=0=V}$), and possible poles of ${p^2H}$ as can be inferred from the curvature invariants, see e.g. \cite{PodolskySamannSteinbauerSvarc2016}.

\section{Geometry of expanding impulses\label{Sec:ImpGeometry}}

\begin{figure}[htb]
\begin{center}
\includegraphics[scale=0.95]{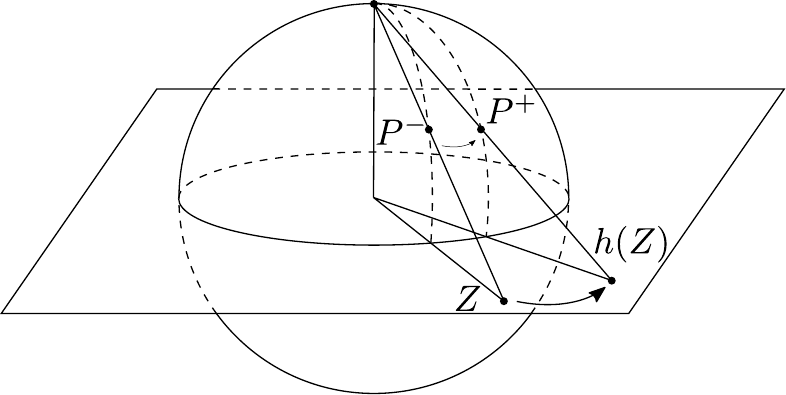}
\caption{\small Stereographic projection corresponds to the mutual identification between points of the Riemann sphere and complex Argand plane. In the case of expanding impulses, it gives direct geometric interpretation to the holomorphic mapping ${h(Z)}$ and the Penrose junction conditions (\ref{JuncCond}).}
\label{SterProj}
\end{center}
\end{figure}

In this section, we will discuss the specific effects of the warp function ${h(Z)}$ representing the source of an expanding impulse. Concerning the Schwarzian derivative (\ref{SchwarzDer}), the wave-like nature of $h(Z)$ can be distinguished. In particular, the general M{\"o}bius transformation of the complex plane to itself,
\begin{equation}
h(Z):\quad Z\ \mapsto \ \frac{a\,Z+b}{c\,Z + d} \,, \label{MobTr}
\end{equation}
leaves $H(Z)$ unchanged and corresponds to the simple Lorentz transformation. On the other hand, going beyond the linear fractional transformation brings the non-trivial Schwarzian derivative. In our case, it will be interpreted as a topological defect related to the presence of cosmic string. To geometrically describe these effects in the flat Minkowski space, it is useful to employ stereographic projection, see e.g. \cite{PodolskyGriffiths00}. The mapping ${Z \mapsto h(Z)}$ in the complex Argand plane corresponds to the geometric identifications of points $P^-$ and $P^+$ on a Riemann sphere, see figure~\ref{SterProj}.
The Riemann sphere can be further identified within the background Cartesian coordinates. In particular, taking the continuous metric (\ref{ContMetric}) and the coordinate transformations leading to the half-spaces \emph{in front of} ($U\geq0$) and \emph{behind} (${U\leq0}$) the impulse, respectively, we find that the impulsive surface ${U=0}$ is a sphere ${(t_\ii^\pm)^2=(x_\ii^\pm)^2+(y_\ii^\pm)^2+(z_\ii^\pm)^2}$, where the $\pm$ sign corresponds to a specific half-space from which $\mathcal{N}$ is approached, and the index $\,_\ii$ indicates values obtained on the impulse~$\mathcal{N}$, i.e., on ${U=0}$. Then, we get
\begin{equation}
Z_\ii=\frac{\eta_\ii^-}{\V_\ii^-}=\frac{x_\ii^-+\im y_\ii^-}{t_\ii^-+z_\ii^-} \,, \qquad h(Z_\ii)=\frac{\eta_\ii^+}{\V_\ii^+}=\frac{x_\ii^++\im y_\ii^+}{t_\ii^++z_\ii^+} \,.
\end{equation}
Inversely, we can write
\begin{equation}
\frac{x_\ii^-}{t_\ii^-}=\frac{Z_\ii+\bar{Z}_\ii}{1+|Z_\ii|^2} \,, \qquad \frac{y_\ii^-}{t_\ii^-}=-\im\frac{Z_\ii-\bar{Z}_\ii}{1+|Z_\ii|^2} \,, \qquad \frac{z_\ii^-}{t_\ii^-}=\frac{1-|Z_\ii|^2}{1+|Z_\ii|^2} \,, \label{StProjM}
\end{equation}
\begin{equation}
\frac{x_\ii^+}{t_\ii^+}=\frac{h+\bar{h}}{1+|h|^2} \,, \qquad \frac{y_\ii^+}{t_\ii^+}=-\im\frac{h-\bar{h}}{1+|h|^2} \,, \qquad \frac{z_\ii^+}{t_\ii^+}=\frac{1-|h|^2}{1+|h|^2} \,, \label{StProjP}
\end{equation}
where the function ${h}$ is evaluated at $Z_\ii$. The expressions (\ref{StProjM}) and (\ref{StProjP}) can thus be understood as the stereographic identification between points in the complex plane and their images on a unit Riemann sphere representing the re-scaled impulsive surface. Subsequently, in terms of such a unit sphere endowed with the Cartesian axes, one can interpret the effects of mapping ${h(Z)}$.

\begin{figure}[htb]
\begin{centering}
\includegraphics[scale=0.95, keepaspectratio]{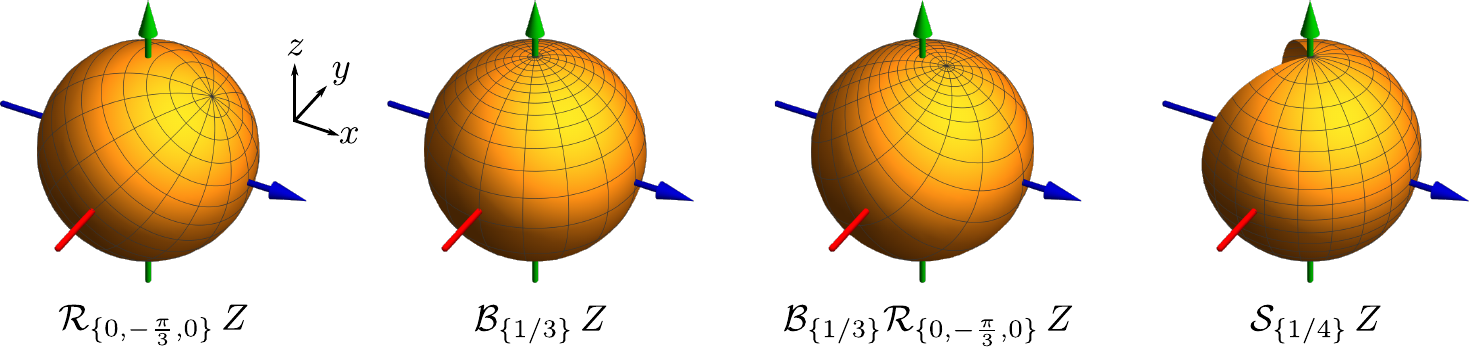}
\caption{Visualisation of specific elementary operations acting on the whole Riemann sphere. Their combinations enter the construction of a particular form of the function $h(Z)$ that, subsequently, encodes the geometric nature of a given expanding impulsive wave described by (\ref{ContMetric}).}\label{elementary_operations}
\end{centering}
\end{figure}

In particular, the construction of explicit form of the function $h(Z)$ can be decomposed into operations representing either pure Lorentz transformations of the form (\ref{MobTr}) or mappings inducing non-trivial Schwartzian derivative (\ref{SchwarzDer}). Here, let us define elementary operations which will be sequentially applied within the following discussion:

\begin{itemize}
    \item  the natural starting point is an \emph{identical} mapping\footnote{The subscript $\,_{j}$ of $h_j(Z)$ identifies particular step in a sequence of the final ${h(Z)}$ construction.}, i.e.,
    \begin{equation}
        h_0(Z)=Z \,. \label{identity}
    \end{equation}
    \item the spatial rotations ${\rot{\varphi,\vartheta,\psi}}$ parameterized by the Euler angles ${\{\varphi,\,\vartheta,\,\psi\}}$ lead to
    \begin{equation}
          h_{j+1}(Z)=\rot{\varphi,\vartheta,\psi} h_j(Z)=e^{\im\varphi}\frac{-\sin(\vartheta/2)+e^{\im\psi}\cos(\vartheta/2)\, h_j}{\cos(\vartheta/2)+e^{\im\psi}\sin(\vartheta/2)\, h_j} \,, \label{rot_param}
    \end{equation}
    see the left part of figure~\ref{elementary_operations} representing the rotated Riemann sphere,
    \item the Lorentz boost in the direction of $z$-axis ${\boost{w}}$, parameterized by the value $w$, gives
    \begin{equation}
        h_{j+1}(Z)=\boost{w} h_j(Z) = w\,h_j(Z) \,,
    \end{equation}
    see the middle plots in figure~\ref{elementary_operations} for a pure boost and its combination with a rotation (\ref{rot_param}),
    \item finally, the simplest string-like structure can be constructed by ``cutting'' out the wedge $2\pi\delta$ around $z$-axis, represented by the action of ${\cstring{\delta}}$, namely
    \begin{equation}
        h_{j+1}(Z)=\cstring{\delta} h_j(Z) = [h_j(Z)]^{1-\delta} \,, \label{cstring}
    \end{equation}
    where the wedge is missing symmetrically in the direction of negative $x$-axis, see the last example in figure~\ref{elementary_operations}. To be more precise, this operation does not only remove a given angle. However, the spherical surface is cut along the ${z-x}$ plane (for the negative $x$ values) and then the angle ${2\pi\delta}$ is opened, while the surface is ``compressed'' (as an accordion or a paper lantern), which may affect already existing defects as we show later.
\end{itemize}
These operations have to be understood as the active transformations of the sphere, while the coordinate system and axes are kept fixed. In general, the Euler angles are arbitrary and it thus seems to be possible to cut out an arbitrary number of strings along different axes, which are moving with different velocities. We will show explicit examples below. Finally, the elementary operations (\ref{identity})--(\ref{cstring}) could be supplemented with other operations, e.g., boosts in the $x$ and $y$ directions. However, these additional operations can be simply understood as their compositions.

\section{Interaction of geodesics with the expanding impulses\label{Sec:Geodesics}}

We aim to analyze properties of particular expanding impulses prescribed by explicit choices of the generating function ${h(Z)}$, see sections~\ref{Sec:OneString} and~\ref{Sec:TwoStrings}. Such a discussion is closely related to the geodesic motion of test observers affected by interaction with the gravitational impulse. However, due to the presence of a kink function in the continuous metric (\ref{ContMetric}), the distributional terms appear in the geodesic equation and its analysis becomes more tricky. The particular case of ${Z=\hbox{const}}$ geodesics was studied in \cite{PodolskySteinbauer2003}. Assuming the ${C^1}$-geodesics, the refraction formulas for their interaction with a generic impulse were derived in \cite{PodolskySvarc2010}. Subsequently, the \emph{existence} and \emph{global uniqueness} of such $C^1$-geodesics crossing expanding impulse, propagating on all constant curvature backgrounds, were rigorously proved in \cite{Steinbauer14, PodolskySamannSteinbauerSvarc2016} using the Filippov solution concept \cite{Filippov:1988, Cortes:2008}.
Such refraction formulas connect initial data ${\{\U_\ii^+,\, \V_\ii^+,\,\eta_\ii^+,\,\dot{\U}_\ii^+,\, \dot{\V}_\ii^+,\,\dot{\eta}_\ii^+\}}$ and ${\{\U_\ii^-,\, \V_\ii^-,\,\eta_\ii^-,\,\dot{\U}_\ii^-,\, \dot{\V}_\ii^-,\,\dot{\eta}_\ii^-\}}$ for the straight lines parameterized by $\tau$, namely
\begin{align}
   \U^\pm =& \dot{\U}_\ii^\pm\,\tau+\U_\ii^\pm \,, \nonumber \\
   \V^\pm =& \dot{\V}_\ii^\pm\,\tau+\V_\ii^\pm \,, \label{flat_geodesics} \\
   \eta^\pm =& \dot{\eta}_\ii^\pm\,\tau+\eta_\ii^\pm \,, \nonumber
\end{align}
i.e., geodesics in the Minkowski half-spaces ${\cal M}^+$ and ${\cal M}^-$, starting/ending on the impulse ${\cal N}$ at ${\tau=0}$, see (\ref{ref:positions}) and (\ref{ref:velocities}) below.

Here, let us summarize the main result of \cite{PodolskySvarc2010, PodolskySamannSteinbauerSvarc2016} important for our further discussion. The explicit $C^1$-matching of geodesics crossing the impulse can be expressed in the form of the refraction formulas encoding the shift of positions and change of the velocities with respect to the fiducial interpretative background. These are derived starting from the fact that the geodesics in coordinates (\ref{ContMetric}) are unique $C^1$-lines across the impulsive wavefront $\mathcal{N}$ given by ${U=0}$, i.e., components of position and velocity evaluated on the impulsive boundary $\mathcal{N}$ (denoted by the subscript $_\ii$) are the same irrespectively whether $\mathcal{N}$ is approached from the region ${\cal M}^+$ with ${U\geq0}$ (denoted by the superscript $^+$) or from the complementary half-space ${\cal M}^-$ with ${U\leq0}$ (denoted by the superscript $^-$). However, to observe the influence of the impulse on test particles, it is natural to employ the fiducial background coordinates (\ref{ExpDoubleNullMink}) for ${\cal M}^+$ and ${\cal M}^-$, respectively. With respect to the background space, the global geodesics do not cross the impulse continuously and the effects of impulse on their motion become explicit. In particular, evaluation of the transformations (\ref{ContTrans+}) and (\ref{NullConeTrans}) on ${U=0}$, and elimination of the continuous coordinates, gives the position shift,
\begin{equation}
\U_\ii^-=|h'|\frac{|Z_\ii|^2}{|h|^2}\,\U_\ii^+ \,, \qquad \V_\ii^-=|h'|\,\V_\ii^+ \,, \qquad \eta_\ii^-=|h'|\frac{Z_\ii}{h}\,\eta_\ii^+ \,, \label{ref:positions}
\end{equation}
while the same procedure for derivatives of (\ref{ContTrans+}) and (\ref{NullConeTrans}) leads to the refraction of the velocities,
\begin{align}
\dot{\U}_\ii^- &= a_{_\U}\dot{\U}_\ii^+ +b_{_\U}\dot{\V}_\ii^+ +\bar{c}_{_\U}\dot{\eta}_\ii^+ +c_{_\U}\dot{\bar{\eta}}_\ii^+ \,, \nonumber \\
\dot{\V}_\ii^- &= a_{_\V}\dot{\U}_\ii^+ +b_{_\V}\dot{\V}_\ii^+ +\bar{c}_{_\V}\dot{\eta}_\ii^+ +c_{_\V}\dot{\bar{\eta}}_\ii^+ \,, \label{ref:velocities} \\
\dot{\eta}_\ii^- &= a_{\eta}\,\dot{\U}_\ii^+ +b_{\eta} \dot{\V}_\ii^+ +\bar{c}_{\eta}\,\dot{\eta}_\ii^+ +c_{\eta}\,\dot{\bar{\eta}}_\ii^+ \,, \nonumber
\end{align}
where the coefficients are constants evaluated on $\mathcal{N}$ with ${Z_\ii}$ obtained via ${h(Z_\ii)=\frac{\eta_\ii^+}{\V_\ii^+}}$, namely
\begin{align}
a_{_\U} &= \frac{1}{|h'|}\left|1+\frac{Z_\ii}{2}\frac{h''}{h'}\right|^2 \,, \\
a_{_\V} &= \frac{1}{4|h'|}\left|\frac{h''}{h'}\right|^2 \,, \\
a_{\eta} &= \frac{1}{2|h'|}\bigg(1 +\frac{Z_\ii}{2}\frac{h''}{h'}\bigg)\frac{\bar{h}''}{\bar{h}'} \,, \\
b_{_\U} &= \frac{|h|^2}{|h'|}\left|1+\frac{Z_\ii}{2}\bigg(\frac{h''}{h'}-2\frac{h'}{h}\bigg) \right|^2 \,, \\
b_{_\V} &= \frac{|h|^2}{4|h'|}\left|\frac{h''}{h'}-2\frac{h'}{h}\right|^2 \,, \\
b_{\eta} &= \frac{|h|^2}{2|h'|}\bigg[1+\frac{Z_\ii}{2}\bigg(\frac{h''}{h'}-2\frac{h'}{h}\bigg)\bigg]\bigg(\frac{\bar{h}''}{\bar{h}'}-2\frac{\bar{h}'}{\bar{h}}\bigg) \,, \\
c_{_\U} &= -\frac{h}{|h'|}\bigg[1+\frac{Z_\ii}{2}\left(\frac{h''}{h'}-2\frac{h'}{h}\right)\bigg]\bigg[1+\frac{\bar{Z}_\ii}{2}\frac{\bar{h}''}{\bar{h}'}\bigg] \,, \\
c_{_\V} &= -\frac{h}{4|h'|}\left(\frac{h''}{h'}-2\frac{h'}{h}\right)\frac{\bar{h}''}{\bar{h}'} \,, \\
c_{\eta} &= -\frac{h}{2|h'|}\bigg[1+\frac{Z_\ii}{2}\left(\frac{h''}{h'}-2\frac{h'}{h}\right) \bigg]\frac{\bar{h}''}{\bar{h}'} \,, \\
\bar{c}_{\eta} &= -\frac{\bar{h}}{2|h'|}\bigg(1+\frac{Z_\ii}{2} \frac{h''}{h'}\bigg)\bigg(\frac{\bar{h}''}{\bar{h}'}-2\frac{\bar{h}'}{\bar{h}}\bigg) \,.
\end{align}
Notice that ${\bar{c}_{_\V} = \overline{c_{_\V}}}$,  ${\bar{c}_{_\U} = \overline{c_{_\U}}}$. As shown in \cite{PodolskySamannSteinbauerSvarc2016}, the normalization of velocity is preserved across the impulse, i.e., ${\dot{\eta}_\ii^-\dot{\bar{\eta}}_\ii^- -\dot{\U}_\ii^-\dot{\V}_\ii^- = \dot{\eta}_\ii^+\dot{\bar{\eta}}_\ii^+ -\dot{\U}_\ii^+\dot{\V}_\ii^+}$. The above (local) expressions do not depend on the Gaussian curvature $\epsilon$, however, to construct a global picture the parameter $\epsilon$ encoding the spacetime foliation has to be considered, see \cite{GriffithsPodolsky:book, PodolskySamannSteinbauerSvarc2016}. The refraction formulas become identical in the trivial case ${h(Z)=Z}$ that implies ${H=0}$ and lacks the impulse. However, one should be careful in the case of non-trivial $h(Z)$ and still trivial ${H=0}$, where the above expressions identify two Minkowski half-spaces via M\"obius transformation (\ref{MobTr}), see section~\ref{Sec:TwoStrings}. This ambiguity in mutual background Cartesian coordinate identification on both sides of the impulse, non-physically affecting in the above refraction formulas, arises from the absence of global Cartesian-like impulsive metric for the expanding waves in contrast to the non-expanding case, see e.g. \cite{PodolskySamannSteinbauerSvarc2015}. 

\begin{figure}[htb]
\begin{center}
\includegraphics[keepaspectratio,scale=0.95]{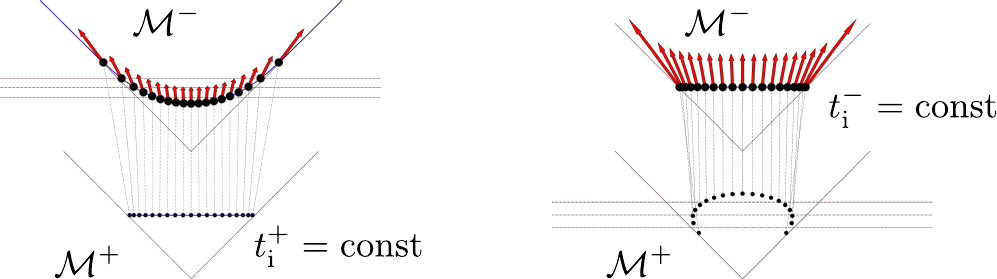}
\caption{A swarm of particles at rest in $\cal{M}^+$ hit by the typical expanding impulse (null cone $\cal{N}$) as seen from the half-space $\cal{M}^+$ (lower part of the schema) and its emergence at null cone $\cal{N}$ as seen from the half-space $\cal{M}^-$ (upper part of the schema). On the left, the test particles interact with the impulse at ${t^+_\ii=\hbox{const}}$ while the right part is the case with ${t^-_\ii=\hbox{const}}$. In this particular case, it can be seen that the particles attain a speed in the radial direction being `attracted' to the axis at the time they leave the wave which causes a formation of caustics later.}
\label{fig:cuts}
\end{center}
\end{figure}

To identify of the pure wave action on the test observers, a specific choice of the initial data ${\{\U_\ii^+,\, \V_\ii^+,\,\eta_\ii^+,\,\dot{\U}_\ii^+,\, \dot{\V}_\ii^+,\,\dot{\eta}_\ii^+\}}$ has to be made which enters the above expressions. Due to the time shift given by a combination of (\ref{ref:positions}) and (\ref{MinkToNull}), we can naturally consider either a swarm of test particles which is hit by the impulsive wavefront ${\cal{N}}$ at the same \emph{constant coordinate time $t^+_\ii$} in $\cal{M}^+$, or vice versa, which appears in the region $\cal{M}^-$ simultaneously at \emph{constant coordinate time $t^-_\ii$}. The second possibility can be also understood as the case of impulse passing through the continuous dust-like distribution of particles where we observe one fixed emergence slice given by constant time $t^-_\ii$. To emphasize the geometric effect of the particular impulse realization encoded in the mapping ${h(Z)}$, we will assume test particles at rest in $\cal{M}^+$ (the rest is defined with respect to the background Minkowskian coordinates, i.e., ${\dot{\U}_\ii^+=\dot{\V}_\ii^+=\dot{\eta}_\ii^+=0}$), which are spherically distributed with radius ${t^\pm_\ii=\hbox{const}}$. A schematic visualization of the above cases (in the simplest one-string situation) is given in figure~\ref{fig:cuts}. The explicit initial data constraints are summarized in the following subsections.

\subsection{Spheres in front of the wave\label{SubSec:Spheres+}}

A Cartesian sphere (formed by test particles) of a constant radius $t_\ii^+$ in front of the wave corresponds to the particle's displacement on a particular cut of the null impulsive cone ${U=0}$ encoded in specific values of the global coordinate $V_\ii$ at the instant of interaction. In particular, evaluation of the transformation (\ref{ContTrans+}) with (\ref{MinkToNull}) on the null cone ${U=0}$ gives the constraint
\begin{equation}
    t_\ii^+= \frac{\cV_\ii}{\sqrt{2}|h'|}\,\frac{1+|h|^2}{1+\epsilon|Z_\ii|^2}=\hbox{const} \,, 
\end{equation}
which, combined with the initial assumption ${t_\ii^+=\hbox{const}}$, fixes the values $V_\ii$ for specific choices of complex plane positions $h(Z_\ii)$ (and inversely $Z_\ii$). The $h(Z_\ii)$ value is related to the Cartesian parameterisation via stereographic projection (\ref{StProjP}),
\begin{equation}
x_\ii^+=\frac{h+\bar{h}}{1+|h|^2}\,t_\ii^+ \,, \qquad
y_\ii^+=-\im\frac{h-\bar{h}}{1+|h|^2}\,t_\ii^+ \,, \qquad
z_\ii^+=\frac{1-|h|^2}{1+|h|^2}\,t_\ii^+ \,, \label{StProjP_t+conct}
\end{equation}
In the case of non-trivial $h(Z)$ with respect to its Schwartzian derivative (\ref{SchwarzDer}), the Cartesian image will suffer for different cutouts. At this moment, it is useful to adopt natural parameterisation,
\begin{equation}
    Z_\ii = \tan\frac{\theta}{2}\,e^{\im\phi} \,, \label{eq:stereo}
\end{equation}
so that ${(x_\ii^\pm,\,y_\ii^\pm,\,z_\ii^\pm)}$ are functions of ${\theta\in\langle 0,\,\pi)}$ and ${\phi\in\langle -\pi,\pi)}$. Subsequently, the deformation of such a test ball is explicitly described by the shift of positions (\ref{ref:positions}), or directly in terms of the continuous coordinates using stereographic projection (\ref{StProjM}) of null cone $\cal{N}$ as viewed from the region behind the impulse. The scale in terms of the original constant radius $t_\ii^+$ is given by
\begin{equation}
   t_\ii^-= \frac{\cV_\ii}{\sqrt{2}}\,\frac{1+|Z_\ii|^2}{1+\epsilon|Z_\ii|^2}=|h'|\,\frac{1+|Z_\ii|^2}{1+|h|^2}\,t_\ii^+ \,, \label{t_i_-_non_const}
\end{equation}
while the deformed surface in front of the wave can be plotted as
\begin{align}
x_\ii^-=|h'|\frac{Z_\ii+\bar{Z}_\ii}{1+|h|^2}\,t_\ii^+ \,, \qquad
y_\ii^-=-\im|h'|\frac{Z_\ii-\bar{Z}_\ii}{1+|h|^2}\,t_\ii^+ \,, \qquad
z_\ii^-=|h'|\frac{1-|Z_\ii|^2}{1+|h|^2}\,t_\ii^+ \, \label{StProjM_t+conct}
\end{align}
which still satisfy ${({x_\ii^-})^2+({y_\ii^-})^2+({z_\ii^-})^2=({t_\ii^-})^2}$, however, ${t_\ii^-}$ is no more a constant.

\subsection{Spheres behind the wave\label{SubSec:Spheres-}}

The second very natural choice of the initial data is such that the test particles form a sphere at a given constant time $t_\ii^-$ behind the wave, i.e., in the region without any strings. The null cone cut is fixed in terms of values $V_\ii$ of the global coordinate $V$ by the condition
\begin{equation}
    t_\ii^-= \frac{\cV_\ii}{\sqrt{2}}\,\frac{1+|Z_\ii|^2}{1+\epsilon|Z_\ii|^2}=\hbox{const} \,, \label{eq:TP}
\end{equation}
and the Cartesian positions on a sphere are related to the values $Z_\ii$ by (\ref{StProjM}),
\begin{equation}
x_\ii^-=\frac{Z_\ii+\bar{Z}_\ii}{1+|Z_\ii|^2}\,t_\ii^- \,, \qquad
y_\ii^-=-\im \frac{Z_\ii-\bar{Z}_\ii}{1+|Z_\ii|^2}\,t_\ii^- \,, \qquad
z_\ii^-=\frac{1-|Z_\ii|^2}{1+|Z_\ii|^2}\,t_\ii^- \,. \label{StProjM_t-conct}
\end{equation}
Viewed from the region ${U>0}$, this corresponds to the deformed initial displacement given by (\ref{StProjP}) satisfying ${({x_\ii^+})^2+({y_\ii^+})^2+({z_\ii^+})^2=({t_\ii^+})^2}$ with non-constant scaling
\begin{equation}
   t_\ii^+= \frac{\cV_\ii}{\sqrt{2}|h'|}\,\frac{1+|h|^2}{1+\epsilon|Z_\ii|^2}=\frac{1}{|h'|}\,\frac{1+|h|^2}{1+|Z_\ii|^2}\,t_\ii^- \,, \label{t_i_+_non_const}
\end{equation}
and Cartesian positions given by
\begin{align}
x_\ii^+=\frac{1}{|h'|}\,\frac{h+\bar{h}}{1+|Z_\ii|^2}\,t_\ii^- \,, \qquad
y_\ii^+=-\frac{\im}{|h'|}\,\frac{h-\bar{h}}{1+|Z_\ii|^2}\,t_\ii^- \,, \qquad
z_\ii^+=\frac{1}{|h'|}\,\frac{1-|h|^2}{1+|Z_\ii|^2}\,t_\ii^- \,. \label{StProjP_t-conct}
\end{align}
The resulting deformed displacement in $\cal{M}^+$ contains defects given by a particular choice of ${h(Z)}$.

\subsection{Visualisations and location of strings}

Although the deformations (\ref{StProjP_t+conct}) and (\ref{StProjP_t-conct}), respectively, do not depend on the value of the Gaussian curvature~$\epsilon$, it is natural to assume the choice $\epsilon=1$. This way, the cuts of the null cone ${U=0}$, ${V_\ii=\mbox{const}}$, parameterised by the remaining global coordinate values $Z_\ii$, are manifestly spheres behind the wave. The radius is $t_\ii^-$ proportional only to ${\cV_\ii}$, namely ${(x_\ii^-)^2+(y_\ii^-)^2+(z_\ii^-)^2=\frac{1}{2}V_\ii^2}$, see (\ref{eq:TP}). This assumption allows for simpler visualisations in terms of the schematic cut and paste figure~\ref{CutAndPAste}, while other choices of $\epsilon$ lead to more complicated sections of the null cone.

To understand the resulting geometry it is natural to investigate where the string ends are attached to the null cone. Their location is related to the scaling factors ${t_\ii^\pm}$. In particular, in the case of initial configuration representing a constant sphere in front of the wave, see section~\ref{SubSec:Spheres+}, these points correspond to the divergence of $t_\ii^-$ given by (\ref{t_i_-_non_const}), while for the constant sphere behind the impulse, see section~\ref{SubSec:Spheres-}, we are inversely looking for zeros of $t_\ii^+$ given by (\ref{t_i_+_non_const}). They thus represent extremes of the `radial' distance of the deformed `spherical' surface.

Determination of the string positions allows to distinguish qualitatively different physical situations. For example, we can arrive at the same string configurations by applying the elementary operations in a different order. However, this typically leads to the different forms of the generating function ${h(Z)}$. Analogously, the rotations will also not change the mutual string configurations, but the functions ${h(Z)}$ will differ. This can be solved by finding the string ends. Therefore, we map the constant sphere-like configuration to obtain its deformed image behind the wave and find points of divergence on such a surface. Then, we can measure mutual angles between all possible pairs of such points for particular functions ${h(Z)}$. Two configurations are identical if angles agree in both cases.

Finally, keep in mind that we infer all the information about the strings from the behaviour on the null cone or properties of its spherical cuts. To proceed more explicitly, the coordinates (\ref{ContMetric+}) in front of the wave, where the strings are present (and, possibly, moving), should be employed in the whole space. However, they are extremely complicated and their analytic inversion to the Cartesian coordinate system, where the topological defects can be directly interpreted, seems to be impossible since it requires finding the inverse of (\ref{ContTrans+}) with (\ref{CoeffContTrans+}).

\section{One string geometries\label{Sec:OneString}}

Although the one-string case has been frequently studied, we would like to provide its description to point out the important technical aspects of the construction that will subsequently appear in more involved two-string cases. The simplest situation of one string located along the $z$ axis, and therefore inducing the deficit angle ${2\pi\delta}$ around it, is given by the mapping
\begin{equation}
    h(Z)=Z^{1-\delta} \,. \label{h_one_string}
\end{equation}

\begin{figure}[H] 
\begin{center}
\includegraphics[keepaspectratio,scale=0.95]{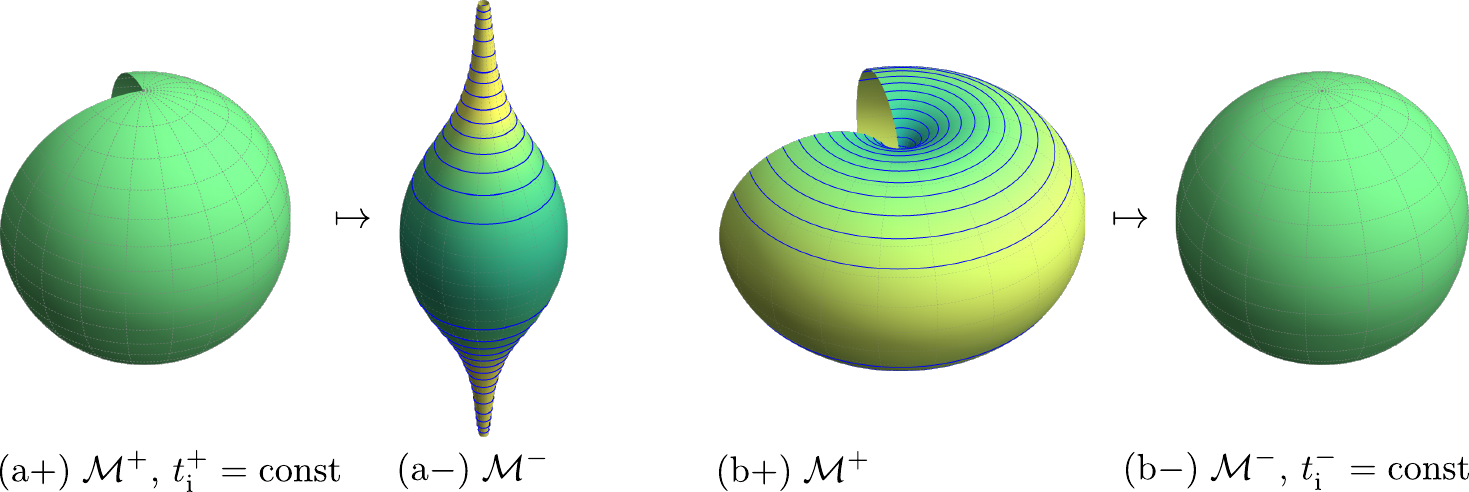}
    \caption{In part (a$+$), the test particles are chosen to form a sphere of a constant radius $t_\ii^+$ and deficit angle ${2\pi\delta}$ along $z$ axis in the region $\mathcal{M}^+$ (in front of the wave). The `sphere' is embedded in the Euclidean space with Cartesian coordinates $(x_\ii^+,y_\ii^+,z_\ii^+)$. The thin dashed lines represent lines of constant $\theta$ and $\phi$ of the stereographic projection (\ref{eq:stereo}). This sphere is distorted behind the wave in $(x_\ii^-,y_\ii^-,z_\ii^-)$ coordinates as visualised in part (a$-$). The particles appear \emph{at different times} and different locations, see (\ref{ref:positions}). The time shift is depicted by a colour change and the lines of constant $t_\ii^-$ are shown as thick blue lines. There is a singularity along the $z^-$ (the spikes reach infinity) while the deficit angle disappears. The part (b$+$) represents such initial conditions in $\mathcal{M}^+$ that the particles emerge as a real geometric sphere in $\mathcal{M}^-$ with radius $t_\ii^-=\hbox{const}$ and coordinates $(x_\ii^-,y_\ii^-,z_\ii^-)$, see part (b$-$).}
    \label{fig:1s-3d}
\end{center}
\end{figure}

\begin{figure}[H]
\begin{center}
\includegraphics[keepaspectratio,scale=0.95]{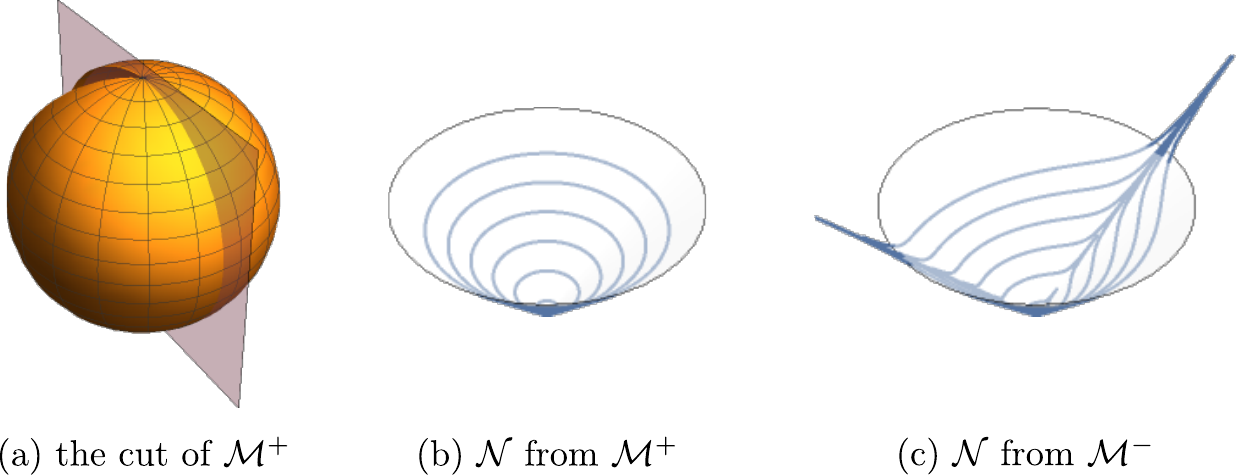}
    \caption{The initial ${t_\ii^+=\hbox{const}}$ sphere is cut by a plane passing through the origin and including both string ends (a). Since the plane does not contain the deficit angle, taking specific values $t_\ii^+$ corresponds to the full circles of test particles (b) that are lying on the null cone $\cal{N}$, where the time goes in the direction of the cone axis. The function ${h(Z)}$ then maps the initial circles (b) along the null cone $\mathcal{N}$ as in figure~\ref{CutAndPAste}. In (c) we can see the resulting displacement of test particles on $\cal{N}$ viewed from the $\mathcal{M}^-$ region. The divergences correspond to the moving string ends related to the infinities of (\ref{t_i_-_non_const}).} \label{NullConeCut_1string}
\end{center}
\end{figure}

As we have already mentioned in section~\ref{Sec:Geodesics}, test particles forming an initial spherical shell in the region $\mathcal{M}^+$, and interacting with the impulse simultaneously at a constant time ${t_\ii^+}$, will be displaced in both space and time directions of $\mathcal{M}^-$, and vice versa. The surfaces representing such initial conditions are visualized in figure~\ref{fig:1s-3d}. In the case of sphere with a constant radius $t_\ii^+$ in $\mathcal{M}^+$, we get 
\begin{equation}
    t_\ii^- \rightarrow \infty \,, \qquad z_\ii^- \rightarrow \pm\infty \,,
\end{equation}
along the axis, corresponding to the (\ref{t_i_-_non_const}) diverges at the places ${Z=\{0,\,\infty\}}$ where the string ends are attached to the null cone, see the left part of the figure~\ref{fig:1s-3d}. This divergence is in reverse translated into the shape of the initial time-slice in the right part of the figure~\ref{fig:1s-3d} leading to particles emerging simultaneously at constant time $t_\ii^-$.
Moreover, to visualize the effect of the Penrose junction condition (\ref{JuncCond}) as a null cone mapping, which was schematically illustrated in figure~\ref{CutAndPAste}, we plot its explicit realization for the one string case (\ref{h_one_string}), see figure~\ref{NullConeCut_1string}.

\begin{figure}[H] 
\begin{center}
\includegraphics[keepaspectratio,scale=0.95]{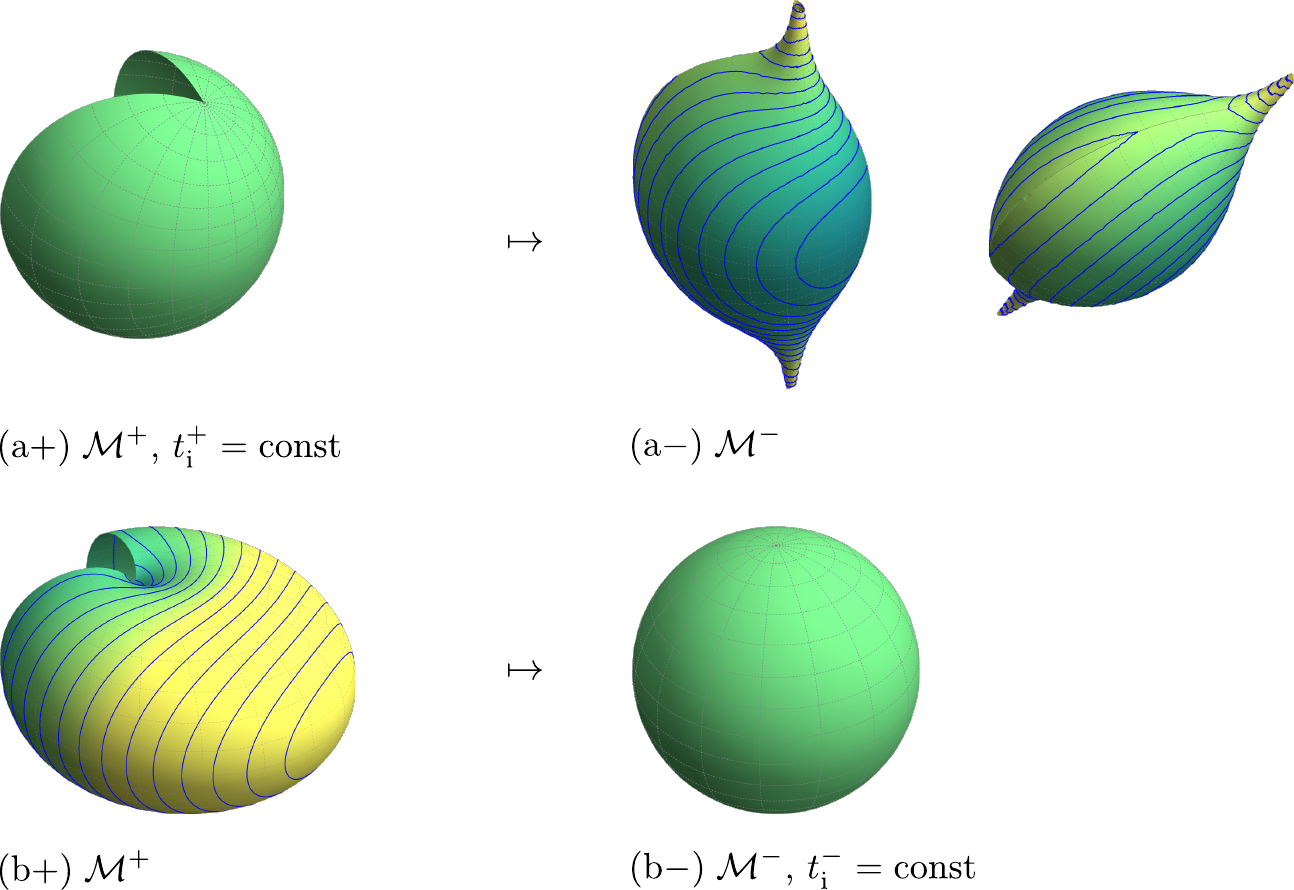}
    \caption{Description of the plot is the same as in figure~\ref{fig:1s-3d}. The differences in the shape of the particles swarm interacting with the impulse are caused by the additional string motion in the $x$ direction related to the form of function $h(Z)$ given by (\ref{one_string_boost}) with the boost parameter ${w=0.6}$. In the second picture in (a$-$) we can observe a sharp `fin' with the time equipotentials being non-smooth. That is the region through which the string travels. In these visualisations, the relation between (a) and (b) subfigures clearly corresponds to the multiplication of the radius by a scaling factor ${|h'|\,\frac{1+|Z_\ii|^2}{1+|h|^2}}$. The string ends are at the north and south pole behind the wave regardless of its transversal motion.}\label{fig:1string-boosted}
\end{center}
\end{figure}

\begin{figure}[H]
\begin{center}
\includegraphics[keepaspectratio,scale=0.95]{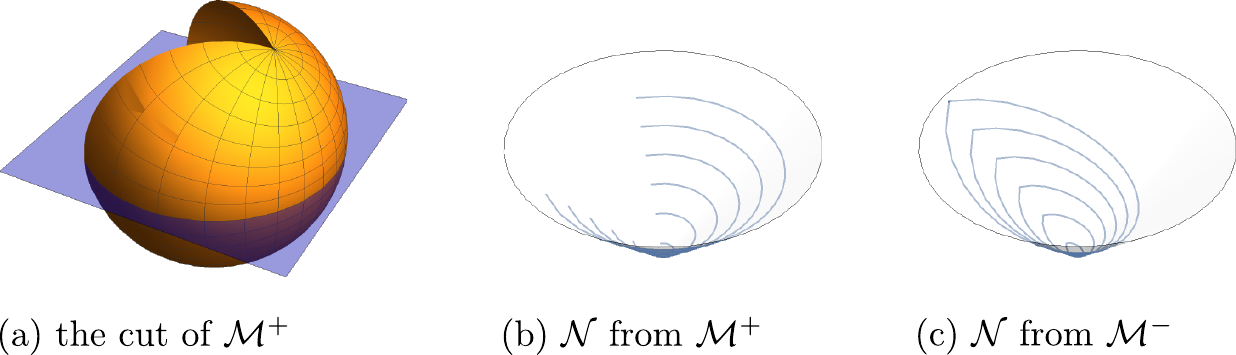}
\caption{The planar cut of the initial swarm of particles is visualized in analogy with  figure~\ref{NullConeCut_1string}. However, here in part (a), the plane is chosen to explicitly contain the topological defect caused by the string (b) while the string ends are out of the plane. The fin-like deformation in part (c) corresponds to the string motion in the $x$ direction, however, it does not contain divergences since the string ends are out of our planar cut.}
\label{fig:1s-boosted-3d}
\end{center}
\end{figure}

The above one-string situation can be non-trivially extended by its boosting in the perpendicular direction. The corresponding complex mapping ${h(Z)}$ can be constructed as a series of string creation along $z$ axis, perpendicular rotation, boost, and final backward rotation, namely
\begin{equation}
h(Z)=\rot{-\frac{\pi}{2},0,0}\boost{w}\rot{\frac{\pi}{2},0,0}\cstring{\delta}Z \,.
\end{equation}
It explicitly becomes
\begin{equation}
h(Z)= -\frac{w-1-(w+1)Z^{1-\delta}}{w+1-(w-1)Z^{1-\delta}}\,. \label{one_string_boost}
\end{equation}
The boost-induced asymmetry reflected in the deformation of the natural static spherical initial conditions, with $t_\ii^+$ or $t_\ii^-$ being constant, respectively, is visualized in figure~\ref{fig:1string-boosted}.
The example of an explicit null cone mapping for the moving string (\ref{one_string_boost}) is visualized in figure~\ref{fig:1s-boosted-3d}.

\begin{figure}[H]
\begin{center}
\includegraphics[keepaspectratio,scale=0.95]{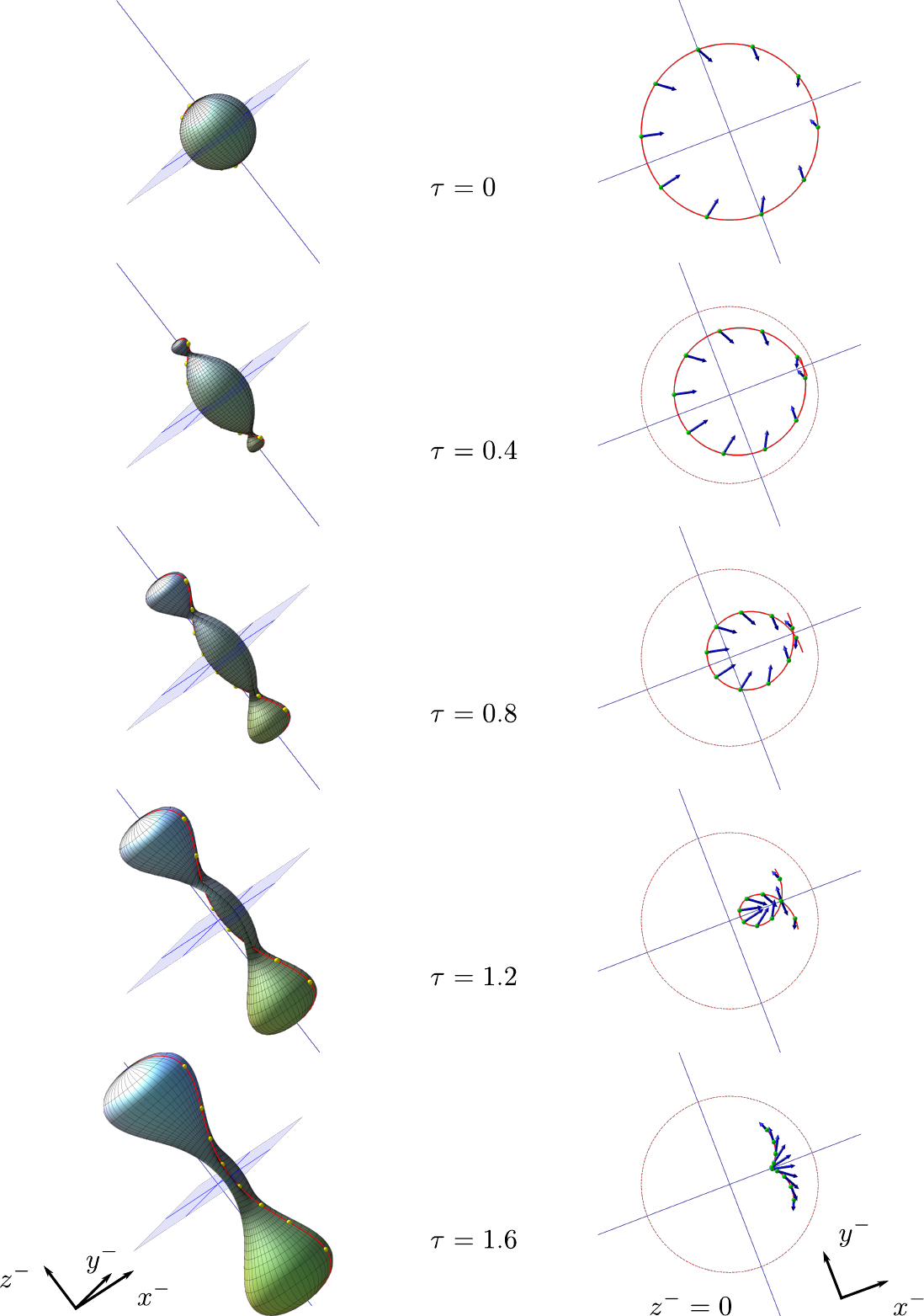}
    \caption{The effect of snapping boosted string (\ref{one_string_boost}) on free test particles initially at rest in $\mathcal{M}^+$ with positions satisfying ${{x_\ii^+}^2+{y_\ii^+}^2+{z_\ii^+}^2={t_\ii^+}^2}$ such that particular choice of ${t_\ii^+}$ leads to ${t_\ii^-=\hbox{const}}$, i.e, the data are prepared as in the bottom part of figure~\ref{fig:1string-boosted} to emerge as a real sphere at given time ${t_\ii^-}$. However, the test observers gain nontrivial velocities. Their time evolution is plotted in the left column. To visualize the asymmetry induced by the string motion in the $x$ direction we show the cut by plane ${z=0}$, perpendicular to the string, in the right column.}\label{one_string_boost_3D}
\end{center}
\end{figure}

Finally, as a comprehensive picture characterizing the dynamical structure of boosted one-string impulsive spacetime with (\ref{one_string_boost}) we plot a sequence of deformations of an initially static swarm of particles. Their initial space and time displacement in the region $\mathcal{M}^+$ is chosen in such a way that they all emerge simultaneously in the region $\mathcal{M}^-$ as a sphere of constant radius ${t_\ii^-}$, i.e, it is described by figure~\ref{fig:1string-boosted} (b${+}$). Subsequently, due to the non-trivial impulsive effect, the particles start moving along geodesics (straight lines) in the Minkowski background (\ref{flat_geodesics}) with initial data given by (\ref{ref:positions}) and (\ref{ref:velocities}), see figure~\ref{one_string_boost_3D}. The vertical deformation and particle acceleration along the $z$ axis are given by their attraction by the moving ends of the snapped string. Near poles, where the string ends are attached to the impulsive sphere, the magnitude of velocity approaches the speed of light. The attractive effect of the string ends results in the caustics formed by a mutual crossing of trajectories starting on opposite sides of the initial sphere with respect to the $z$ axis. Moreover, the horizontal asymmetry is induced by the boost of the string in the $x$ direction. To emphasize the boost effect we plot the cut by ${z=0}$ plane in the left part of figure~\ref{one_string_boost_3D}.

The above discussion shows the effects of a moving snapped cosmic string on free test observers and the way how these effects can be understood and visualized. Notice that the subcase without string boost was already studied in \cite{PodolskySteinbauer2003, PodolskySvarc2010} considering also cosmological backgrounds. In the subsequent section, we follow \cite{NutkuPenrose92, PodolskyGriffiths00} and analyze various impulses generated by a pair of snapped strings in the same way as in the one-string case. 

Naturally, in the case of boosted string one can be worried by the fact that the cut out planes are not extrinsically flat, as can be seen in figure \ref{fig:1string-boosted}. However, it can be analytically shown that following situations are equivalent:
\begin{itemize}
\item initially static particles and their interaction with impulse generated by a string boosted by $w$ in the positive $x$ direction, i.e., ${h(Z)}$ given by (\ref{one_string_boost}),
\item initially moving particles in the negative $x$ direction with the velocity magnitude $v=\frac{1-w^2}{1+w^2}$ interacting with a static string described by (\ref{h_one_string}).
\end{itemize}

\section{Two strings geometries\label{Sec:TwoStrings}}

Here, we review and extend the results presented in \cite{PodolskyGriffiths00} and discuss their geometric properties via analysis of induced geodesic motion. Subsequently, an alternative sequences of basic steps entering the two strings complex mapping $h{(Z)}$ construction will be employed and particular differences in the resulting motion identified.

\subsection{Original results by Podolsk\'y and Griffiths\label{SubSec:PGOriginal}}

The possibility of expanding impulse generated by a colliding and snapping pair of cosmic strings was originally anticipated by Nutku and Penrose in \cite{NutkuPenrose92}. Simultaneously, there were doubts that explicit realization of the corresponding function $h{(Z)}$ is hard to find, however, its existence should be guaranteed by the Riemann theorem.  Surprisingly, a few years later Podolsk{\'y} and Griffiths explicitly performed such a construction in \cite{PodolskyGriffiths00}. Their simplest non-boosted two-string formula reads
\begin{equation}
h(Z)=\cstring{\varepsilon}\rot{\frac{\pi}{2},\frac{\pi}{2},\frac{\pi}{2}}\cstring{\delta}= \left(\frac{iZ^{1-\delta}-1}{Z^{1-\delta}-i}\right)^{1-\varepsilon} \,. \label{PG2Strings_fce_h}
\end{equation}
In terms of fixed Cartesian coordinates, it can be described in such a way that the string parameterized by $\delta$ and placed along the $z$ axis is rotated to take a place along the $y$ axis and the second string (encoded in $\varepsilon$) is then created along the $z$ axis. However, the construction should be more precisely understood in terms of the active Lorentz transformations (\ref{MobTr}). This can be seen from the no string limit of (\ref{PG2Strings_fce_h}). Taking both parameters trivial, namely ${\delta=0=\varepsilon}$, does not provide the identity, but 
\begin{equation}
h(Z)= \frac{iZ-1}{Z-i} \,,
\end{equation}
which is exactly the residual rotation ${\rot{\frac{\pi}{2},\frac{\pi}{2},\frac{\pi}{2}}\,Z}$ employed within construction of (\ref{PG2Strings_fce_h}). The Schwarzian derivative (\ref{SchwarzDer}) is vanishing and the seemingly non-trivial position shift (\ref{ref:positions}) directly shows the unphysical rotation or the artificial identification of the background Cartesian frame. However, to directly gain all relevant information about test particles interacting with the impulse from the refraction formulas (\ref{ref:positions}) and (\ref{ref:velocities}), it is important and useful to remove such a coordinate discrepancy. Before we do that, let us show the ultimate result of \cite{PodolskyGriffiths00} adding a boost to the simplest interaction of static strings (\ref{PG2Strings_fce_h}), and therefore, interpreted as the collision of strings that induces their snap and subsequent creation of the impulse. The particular complex mapping reads
\begin{equation}
    h(Z)= w_2 \frac{h_c^{1-\varepsilon}-1}{h_c^{1-\varepsilon}+1} \label{PG2Strings_fce_h_boost}
\end{equation}
with
\begin{equation}
    h_c(Z)= -\frac{(w_1-i)Z^{1-\delta}+(w_1+i)}{(w_1+i)Z^{1-\delta}+(w_1-i)} \,.
\end{equation}

\begin{figure}[htb]
\begin{center}
\includegraphics[keepaspectratio,scale=0.95]{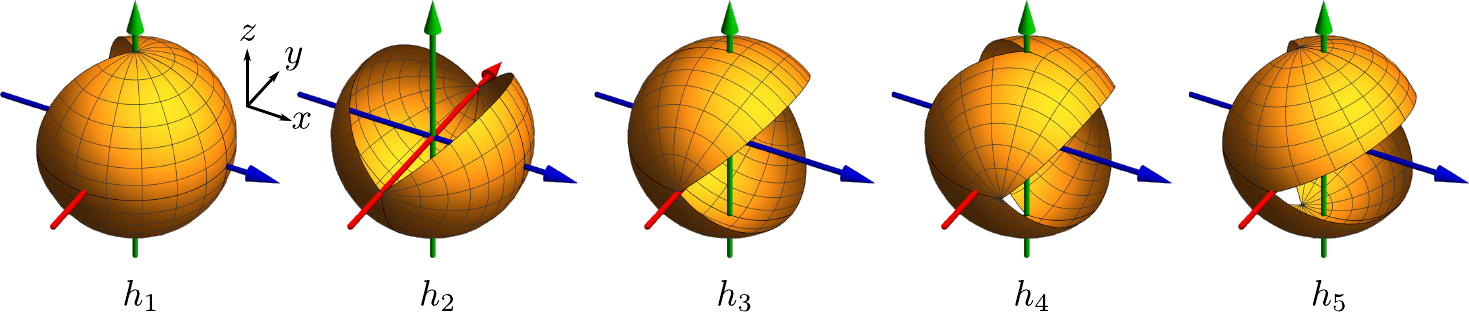}
\end{center}
\caption{A sequence of elementary mappings (\ref{PG2Strings_operations_h0})--(\ref{PG2Strings_operations_h5}) employed within the construction of the function (\ref{PG2Strings_fce_h_boost_improved}) encoding interaction of two cosmic strings and their snap. Here we plot the simplest case $w_2=w_3=w_5=1$ directly improving (\ref{PG2Strings_fce_h}). The function $h(Z)$ then enters the junction conditions (\ref{JuncCond}) of two half-spaces `in front of' and `behind' the impulsive wave. Interestingly, the second string creation, step ${h_3\,\rightarrow\,h_4}$, inherently induces an additional boost. This can be counterbalanced by a specific choice of $w_j$, see figure~\ref{PG2Strings_fce_h_boost_gen_Riemann_spec_boost}.} \label{PG2Strings_fce_h_boost_gen_Riemann_triv_boost}
\end{figure}

\begin{figure}[htb]
\begin{center}
\includegraphics[keepaspectratio,scale=0.95]{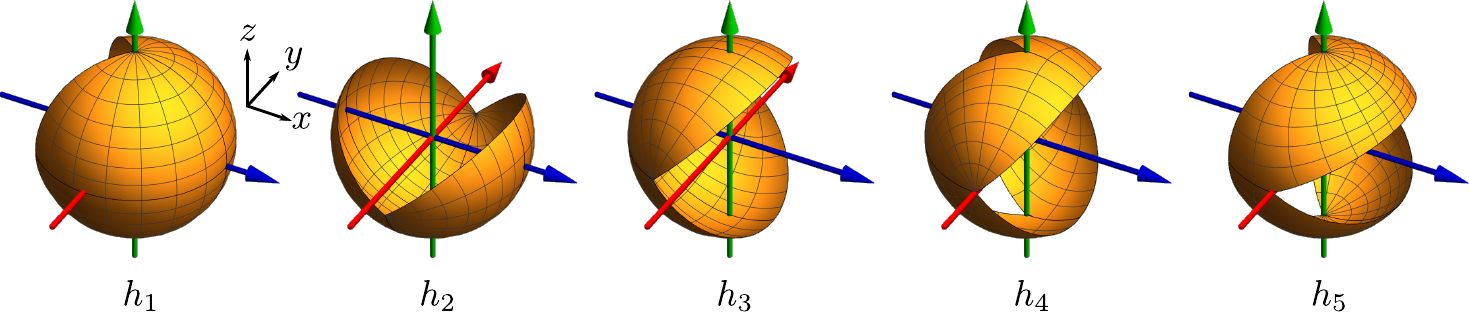}
\end{center}
\caption{The same sequence of elementary mappings as in the figure~\ref{PG2Strings_fce_h_boost_gen_Riemann_triv_boost}. However, here we decided to take ${w_2=\tan\frac{\pi}{4(1-\varepsilon)}}$ and ${w_3=1=w_5}$ so that the additional inherent boost in ${h_3\,\rightarrow\,h_4}$ is compensated and the strings are placed along the axes, compare $h_4$ steps.} \label{PG2Strings_fce_h_boost_gen_Riemann_spec_boost}
\end{figure}

Now, to improve the artificial coordinate effect in (\ref{PG2Strings_fce_h}), and subsequently also in (\ref{PG2Strings_fce_h_boost}), let us perform the following sequence of mappings, namely
\begin{align}
h_0(Z) &= Z \,, \label{PG2Strings_operations_h0} \\
h_1(Z) &= \cstring{\delta}\,h_0(Z) \,, \\
h_2(Z) &= \boost{w_2}\rot{\frac{\pi}{2},\frac{\pi}{2},\pi}\,h_1(Z) \\ 
h_3(Z) &= \boost{w_3}\rot{-\frac{\pi}{2},0,0}\,h_2(Z) \,,\\
h_4(Z) &= \cstring{\varepsilon}\,h_3(Z) \,, \label{PG2Strings_operations_h4}\\
h_5(Z) &= \boost{w_5}\rot{\frac{\pi}{2},\frac{\pi}{2},\frac{\pi}{2}}\,h_4(Z) \,. \label{PG2Strings_operations_h5}
\end{align}
These operations can be geometrically understood in terms of the Riemann sphere, see figures~\ref{PG2Strings_fce_h_boost_gen_Riemann_triv_boost} and~\ref{PG2Strings_fce_h_boost_gen_Riemann_spec_boost} for the special cases, while the fully general mapping takes the form
\begin{equation}
    h(Z)= i w_5 
    \frac{-1+i\left(w_3\frac{(i+w_2)Z^\delta-(i-w_2)Z}{(i-w_2)Z^\delta-(i+w_2)Z}\right)^{1-\varepsilon}}
    {1+i\left(w_3\frac{(i+w_2)Z^\delta-(i-w_2)Z}{(i-w_2)Z^\delta-(i+w_2)Z}\right)^{1-\varepsilon}}
    \,, \label{PG2Strings_fce_h_boost_improved}
\end{equation}
which naturally becomes identical for trivial boosts and deficit angles, i.e., ${w_j=1}$ and ${\delta=0=\varepsilon}$. The simplest static interaction of the strings is then described by choosing $w_2=w_3=w_5=1$ in (\ref{PG2Strings_fce_h_boost_improved}) which then corresponds to (\ref{PG2Strings_fce_h}) with the artificial coordinate rotation removed, namely
\begin{equation}
    h(Z)=i\frac{-1+i\left(-i\frac{1-iZ^{1-\delta}}{1+iZ^{1-\delta}}\right)^{1-\varepsilon}}{1+i\left(-i\frac{1-iZ^{1-\delta}}{1+iZ^{1-\delta}}\right)^{1-\varepsilon}}     \,, \label{PG2Strings_fce_h_improved}
\end{equation}
while for a specific choice of boosts ${w_j}$ we may obtain an improved version of (\ref{PG2Strings_fce_h_boost}). On a general level, the location of ends of the strings for (\ref{PG2Strings_fce_h_boost_improved})  is given by
\begin{equation}
Z=\left\{
0,
\infty,
\left(\frac{i+w_2}{i-w_2}\right)^{\frac{1}{1-\delta}},
\left(\frac{i-w_2}{i+w_2}\right)^{\frac{1}{1-\delta}}
\right\} \,. \label{PG2Strings_fce_h_boost_improved_zeros}
\end{equation}

\begin{figure}[H]
\begin{center}
\includegraphics[keepaspectratio,scale=0.95]{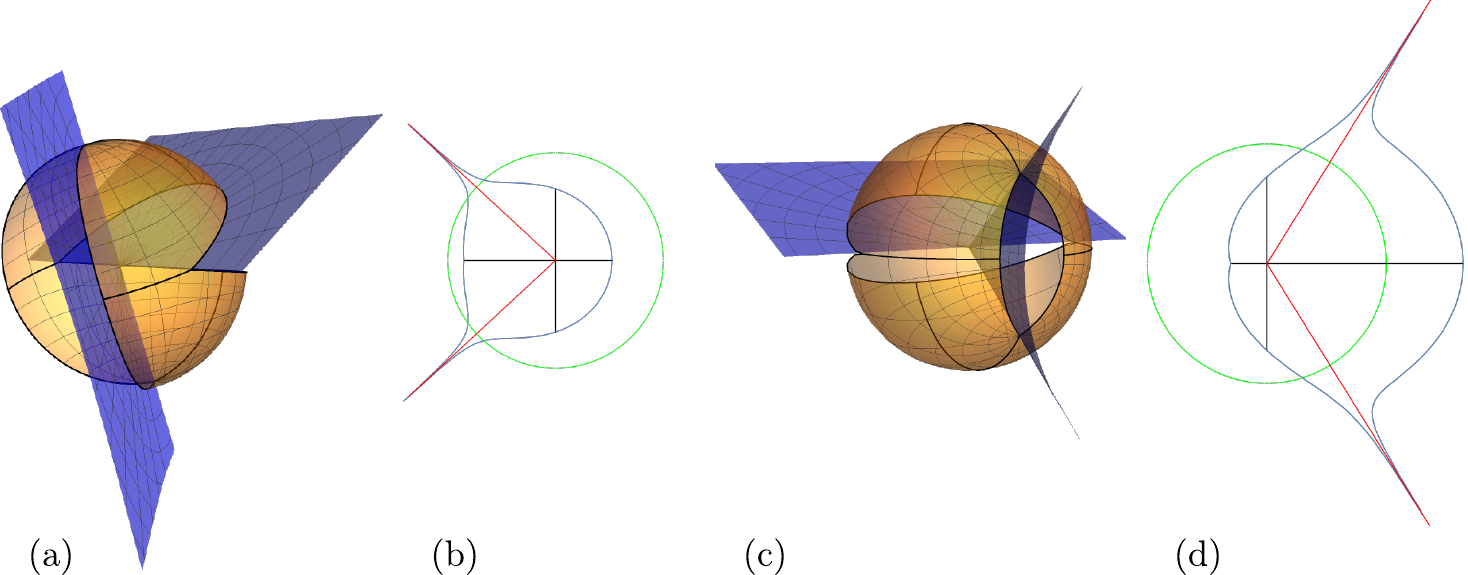}
\caption{In part (a) we plot the Riemann sphere as in figure \ref{PG2Strings_fce_h_boost_gen_Riemann_spec_boost} $h_5$, i.e., corresponding to the mapping (\ref{PG2Strings_fce_h_boost_improved}) with ${w_2=\tan\frac{\pi}{4(1-\epsilon)}}$ compensating the additional boost. The cut by $x$-$y$ plane of the $t_\ii^+=\mbox{const}$ initial condition image in $\mathcal{M}^-$ is shown in part (b). The string ends follow the red lines, whose mutual angle is given by (\ref{eq:angles}). The same situations with a generic boost ${w_2=0.4}$ are visualized in figures (c) and (d). In (a) and (c) we plot the blue surfaces given by parameterization of the wedge edges. If there is no transversal motion of the strings, the surfaces are planes with vanishing extrinsic curvature, while in the case of boosted strings, the situation becomes more complicated.}\label{PG2Strings_velocity_comparison}
\end{center}
\end{figure}

Here, let us also emphasize that the elementary operation $\cstring{\varepsilon}$, see (\ref{PG2Strings_operations_h4}), which creates the second string placed along $z$-axis, inherently shifts the position in the $x$-direction of the already existing string. In figure~\ref{PG2Strings_fce_h_boost_gen_Riemann_triv_boost}, this corresponds to the step $h_3\,\rightarrow\,h_4$, where the sphere is distorted and the existing string is effectively boosted in the $x$-direction. We can counterbalance this effect by additional boost in the opposite direction, namely
\begin{equation}
    w_2 = \tan\frac{\pi}{4(1-\varepsilon)} \,, \label{PG2Strings_fce_h_boost_improved_spec_boost_value}
\end{equation}
and ${w_3=1=w_5}$, see figure~\ref{PG2Strings_fce_h_boost_gen_Riemann_spec_boost} and changes in the sequence $h_2\,\rightarrow\,h_3\,\rightarrow\,h_4$. The resulting nodal points are then aligned with the axes and there is no transversal velocity (as could be seen from their action on test particles). These situations are also compared in figure~\ref{PG2Strings_velocity_comparison}.

This discussion is connected with the real location of the strings viewed from the region behind the wave. Geometrically, after cutting out two wedges (deficit angles) we are glueing the corresponding `lips' back together, however, it is done in a particular order. Here, the string governed by parameter $\delta$ remains straight along the $z$ axis, while using (\ref{PG2Strings_fce_h_boost_improved_zeros}) the string parts governed by $\varepsilon$, and lying in the ${x}$-${y}$ plane, can be identified by the polar angle $\phi_s$, namely
\begin{equation}
    \phi_s = \pm \frac{\pi-\arg\frac{w_2+i}{w_2-i}}{1-\delta} \,. \label{eq:angles}
\end{equation}
In the case with (\ref{PG2Strings_fce_h_boost_improved_spec_boost_value}) and ${w_3=1=w_5}$, i.e., the constants are set so that there is no transversal motion of the strings, pieces of one string are attached to the north and south pole, respectively, while the second string ends form the mutual angle
\begin{equation}
    \Delta\phi=\left(2-\frac{1}{2(1-\delta)(1-\varepsilon)}\right)\pi \,,
\end{equation}
which is found as a direct application of the general formula (\ref{eq:angles}).


\begin{figure}[H] 
\begin{center}
\includegraphics[keepaspectratio,scale=0.95]{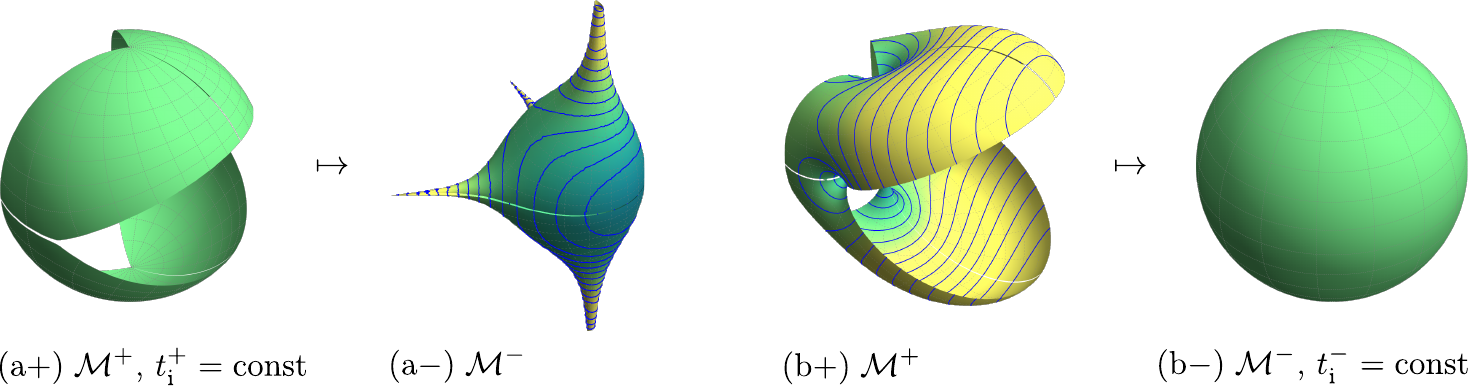}
\caption{As in figure~\ref{fig:1s-3d} two natural sets of static initial data are prepared for the improved two-string case (\ref{PG2Strings_fce_h_boost_improved}) with ${w_3=1=w_5}$ and $w_2$ given by (\ref{PG2Strings_fce_h_boost_improved_spec_boost_value}). The plot of the initial condition for (\ref{PG2Strings_fce_h}) would differs by rotation and a trivial choice $w_2=1$.
}\label{fig:2strings_PG_imp_initial_conditions}
\end{center}
\end{figure}

\begin{figure}[H]
\begin{center}
\includegraphics[keepaspectratio,scale=0.95]{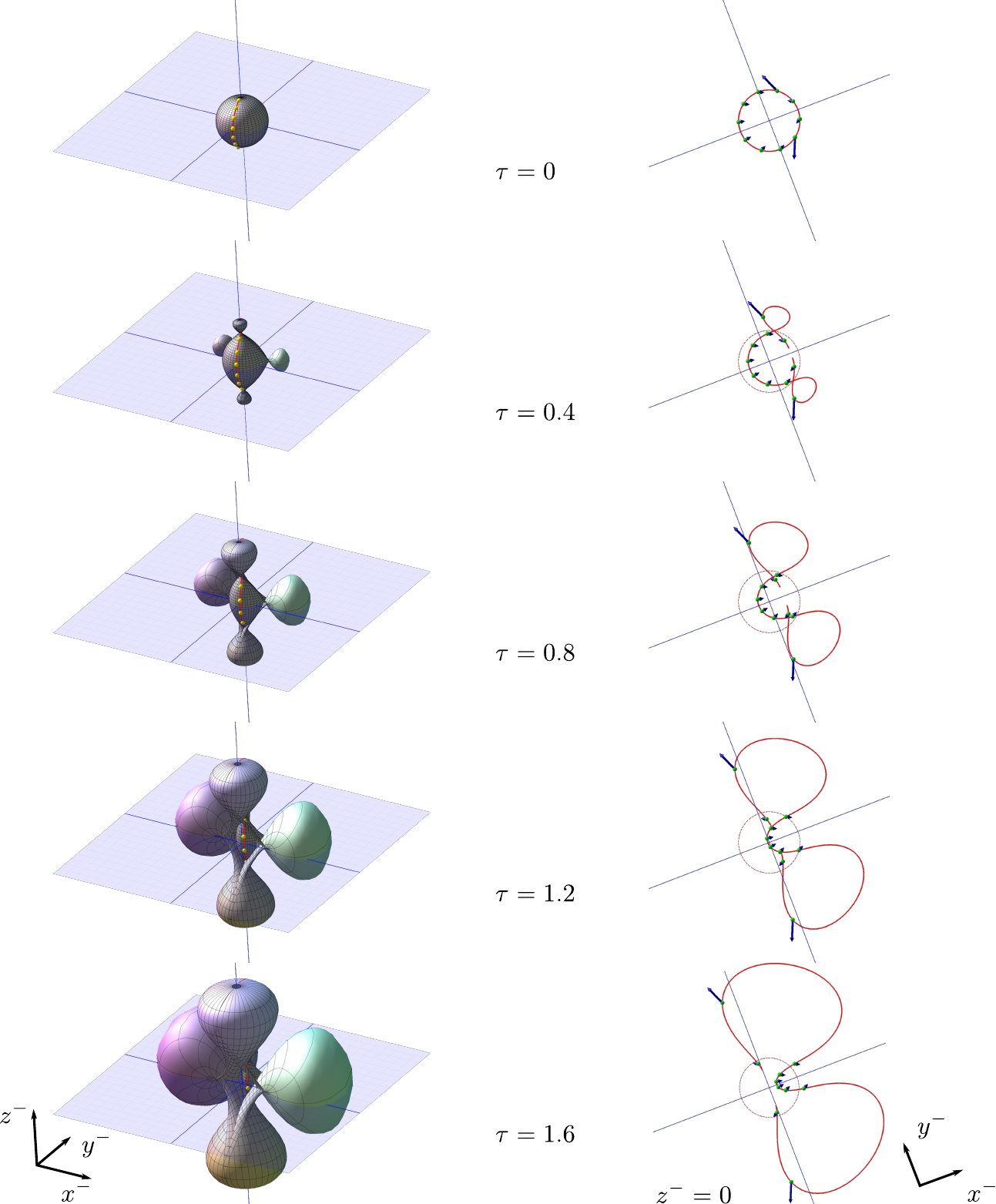}
    \caption{In analogy with figure~\ref{one_string_boost_3D} we plot a burst of test particles hit by the impulsive wave generated by the improved mapping (\ref{PG2Strings_fce_h_improved}), which can be directly used in refraction formulas (\ref{ref:positions}) and (\ref{ref:velocities}) entering geodesics (\ref{flat_geodesics}). This choice is visualized in figure~\ref{PG2Strings_fce_h_boost_gen_Riemann_triv_boost}. The employed initial conditions represent static particles in front of the wave displaced to emerge synchronously at constant time $t_\ii^-$. The typical values ${\delta=\frac{1}{4}}$ and ${\varepsilon=\frac{1}{8}}$ are used. We observe the dragging of test particles in the directions of four moving ends of the snapped string pairs. This is combined with inherent (non-compensated) boost in the $x$ direction induced by the second string creation.}\label{2strings_PG_3D}
\end{center}
\end{figure}

\begin{figure}[H]
\begin{center}
\includegraphics[keepaspectratio,scale=0.85]{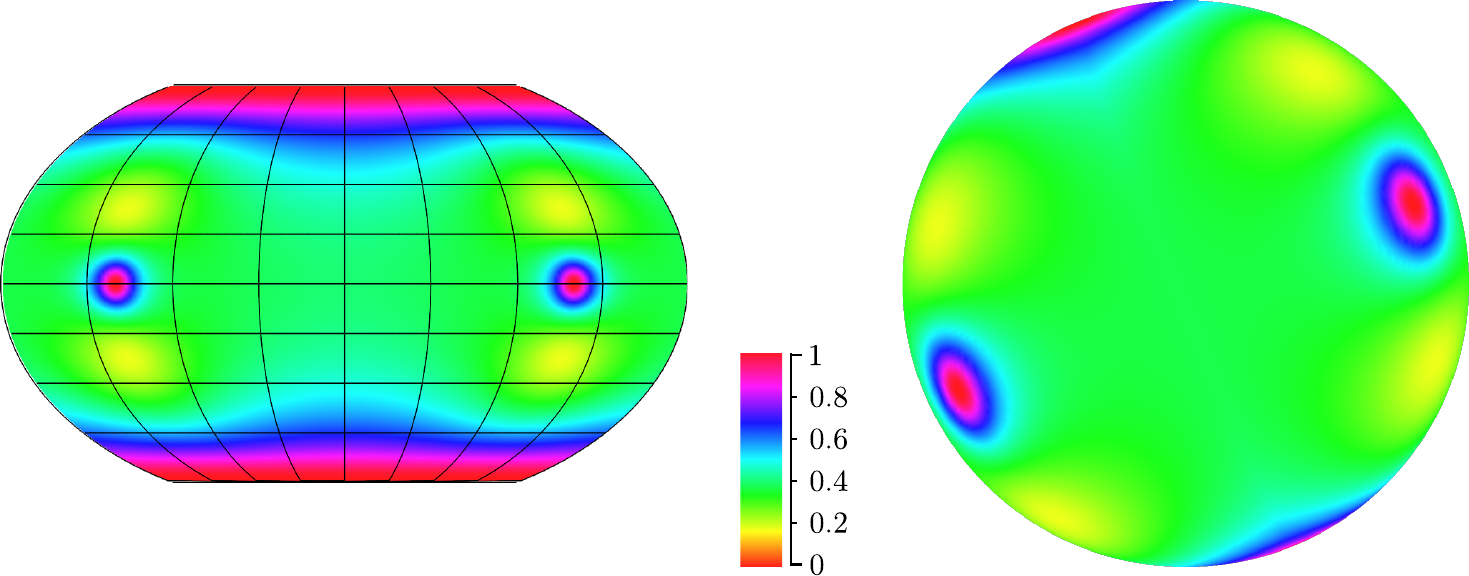}
\caption{The distribution of velocity magnitudes (\ref{velocity_magnitude}) in the region behind the impulse. The initially static particles in front of the impulse emerging simultaneously behind the wave generated by (\ref{PG2Strings_fce_h_improved}) are considered with the typical values ${\delta=\frac{1}{4}}$ and ${\varepsilon=\frac{1}{8}}$ of the deficit parameters. The magnitude reaches the speed of light near the four string ends. Here, we show the Kavrayskiy VII projection (left) and the spatial scheme (right).}
\label{fig:2st_velocities}
\end{center}
\end{figure}

Finally, based on the geodesic motion let us describe the effect of an impulse generated by (\ref{PG2Strings_fce_h_improved}). We visualize its interaction with initially static test particles prepared in $\mathcal{M}^+$ to emerge synchronously on a sphere in $\mathcal{M}^-$, see figure~\ref{fig:2strings_PG_imp_initial_conditions} identifying the initial data and figure~\ref{2strings_PG_3D} depicting the overall deformation caused by the impulse. The shape evolution corresponds to the straight motion on a flat background. However, there is a non-trivial distribution of velocities given by (\ref{ref:velocities}). The map of velocity magnitude,
\begin{equation}
    v^-=\sqrt{(v_x^-)^2+(v_y^-)^2+(v_y^-)^2} \label{velocity_magnitude}
\end{equation}
with
\begin{equation}
    \left(v_x^-,\,v_y^-,\,v_z^-\right)-\equiv \left(\frac{\dot{x}_\ii^-}{\dot{t}_\ii^-},\,\frac{\dot{y}_\ii^-}{\dot{t}_\ii^-},\,\frac{\dot{z}_\ii^-}{\dot{t}_\ii^-}\right)  \,,
\end{equation}
is plotted in figure~\ref{fig:2st_velocities}, which again indicates the string ends since in their neighbourhood the test particles approach the speed of light.

\subsection{Peculiar form of the simple two-string mapping}

In this part, let us show an alternative construction of the function $h(Z)$ in the case of a pair of cosmic strings, which should demonstrate the subtlety of the construction process.

\begin{figure}[H]
\begin{center}
\includegraphics[keepaspectratio,scale=0.95]{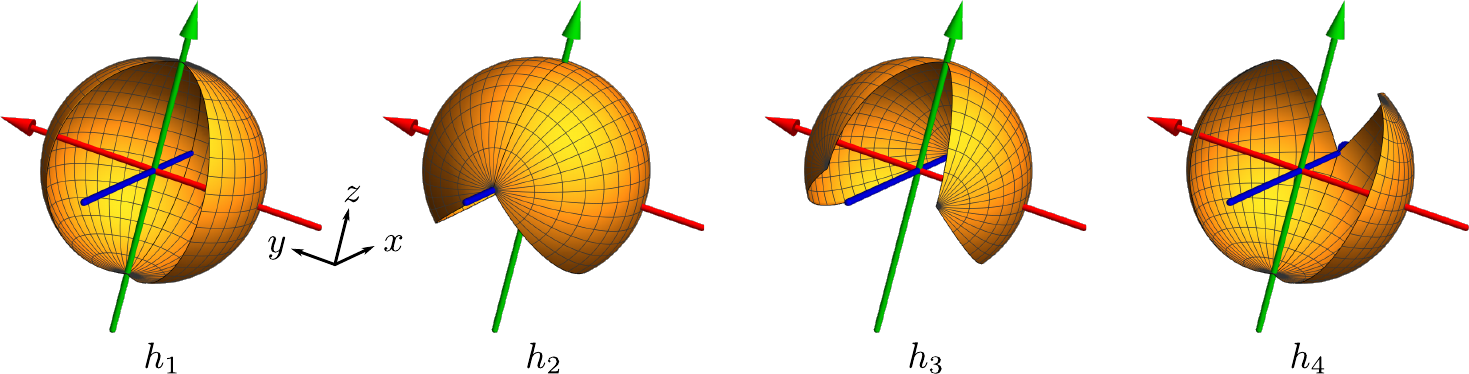}
\caption{Particular steps of the construction of peculiar mapping (\ref{KKS_fce_h_general}) simplified by assuming trivial boost parameters ${w_3=1=w_7}$.}\label{KKS_construction}
\end{center}
\end{figure}

Let us start with the sequence of elementary steps
\begin{align}
h_0(Z) &= Z \,, \label{2strings_pec_h0} \\
h_1(Z) &= \boost{w_3}\cstring{\delta}\,h_0(Z) \,, \\
h_2(Z) &= \rot{\frac{\pi}{2},0,0}\,h_1(Z) \,,\\
h_3(Z) &= \boost{w_7}\cstring{\varepsilon}\,h_2(Z) \,, \label{KKS_2nd_string_creation} \\ 
h_4(Z) &= \rot{-\frac{\pi}{2},0,0}\,h_3(Z) \label{2strings_pec_h4} \,,
\end{align}
which leads to the mapping
\begin{equation}
    h(Z)=-\frac{1-w_7\left(\frac{1+w_3 Z^{1-\delta}}{1-w_3 Z^{1-\delta}}\right)^{1-\epsilon}}%
    {1+w_7\left(\frac{1+w_3 Z^{1-\delta}}{1-w_3 Z^{1-\delta}}\right)^{1-\epsilon}} \,. \label{KKS_fce_h_general}
\end{equation}
Surprisingly, we identify only three string ends located at
\begin{equation}
    Z=\left\{0,\,\infty,\,w_3^{-\frac{1}{1-\delta}}\right\} \,. \label{Pec_St_positions}
\end{equation}
 The sequence (\ref{2strings_pec_h0})--(\ref{2strings_pec_h0}) thus cuts out the wedge $2\pi\delta$ along $z$ axis and performs a boost, makes rotation about $y$ axis, applies another string-like cut parameterized by $\varepsilon$ along the original $x$ axis with another boost, and finally rotates backwards about unchanged $y$ axis. The simplified construction is visualized in figure \ref{KKS_construction}. The mapping then leads to identity for ${\delta=0=\varepsilon}$ and ${w_3=1=w_7}$.

\begin{figure}[H]
\begin{center}
\includegraphics[keepaspectratio,scale=0.95]{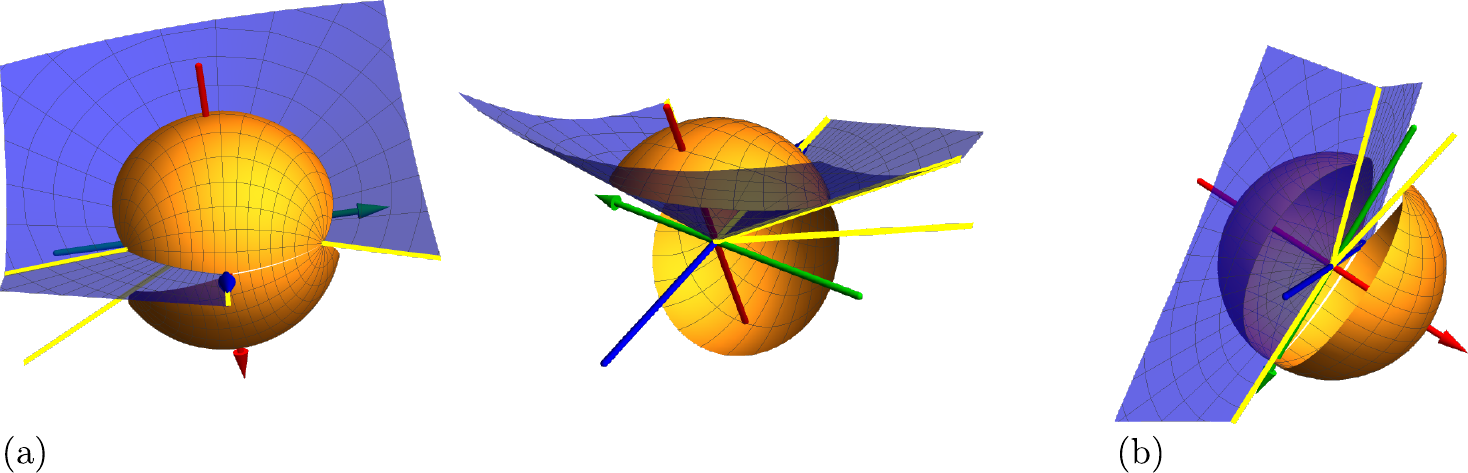}
\caption{
In the case of generic boosts, one surface swept by the Riemann sphere edge has non-vanishing extrinsic curvature, while the second one is a trivial plane, see two views in part (a). In the static case, given by the boost choice (\ref{KKS_boost_static}), all the surfaces are planar. The pictures are reflectively symmetric with respect to the ${y=0}$ plane corresponding to the blue and green axis .}\label{KKS_velocity_comparison}
\end{center}
\end{figure}

Similarly to in the previous section and the case of mapping (\ref{PG2Strings_fce_h_boost_improved}), the second string creation induces an additional boost. Then the edges of Riemann sphere cutouts, parameterized by $V$, represent a generic curved surface. The corresponding extrinsic curvature vanishes and the surface becomes a plane if the inherent boost is compensated by a suitable choice of the artificial boosts in our construction. In particular, taking
\begin{align}
    w_3&=1\,, &
    w_7&=\sqrt{\frac{1-\sin\frac{\epsilon\pi}{2}\left(\tan\frac{\delta\pi}{2}\right)^{-(1-\epsilon)}}%
    {1-\sin\frac{\varepsilon\pi}{2}\left(\tan\frac{\delta\pi}{2}\right)^{+(1-\varepsilon)}}} \label{KKS_boost_static}
\end{align}
guarantees that the strings do not move in the transverse direction at all. This is illustrated in figure~\ref{KKS_velocity_comparison}.

\begin{figure}[H]
\begin{center}
\includegraphics[keepaspectratio,scale=0.95]{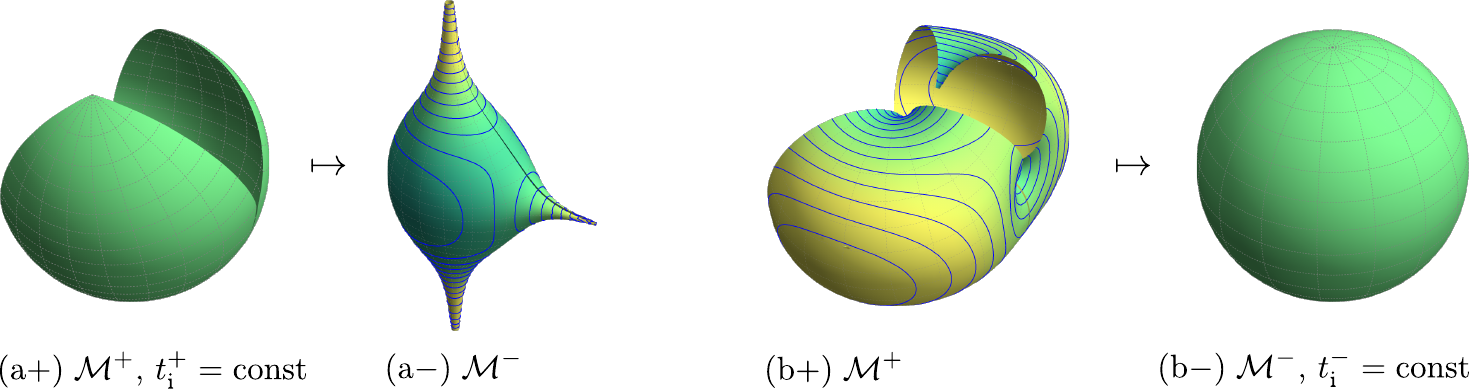}
    \caption{The static initial data choice for the simplified complex mapping (\ref{KKS_fce_h_general}) with ${w_3=1=w_7}$. The infinities in the part (a$-$) indicate the presence of only three moving string pieces.}\label{KKS_data}
\end{center}
\end{figure}

Based on the construction with two string-like operations, one would expect that it describes a pair of perpendicular cosmic strings with four moving ends after their snap. However, inspecting (\ref{Pec_St_positions}) related to the static initial data choice in figure~\ref{KKS_data}, we may conclude that there are only \emph{three} string pieces. This is exactly the above-mentioned subtlety in the ${h(Z)}$ construction. One part of the first string disappears by the creation of the second string (\ref{KKS_2nd_string_creation}) which removes the associated nodal point of the Riemann sphere, see ${h_2\,\rightarrow\,h_3}$ in figure~\ref{KKS_construction}.

\begin{figure}[H]
\begin{center}
\includegraphics[keepaspectratio,scale=0.95]{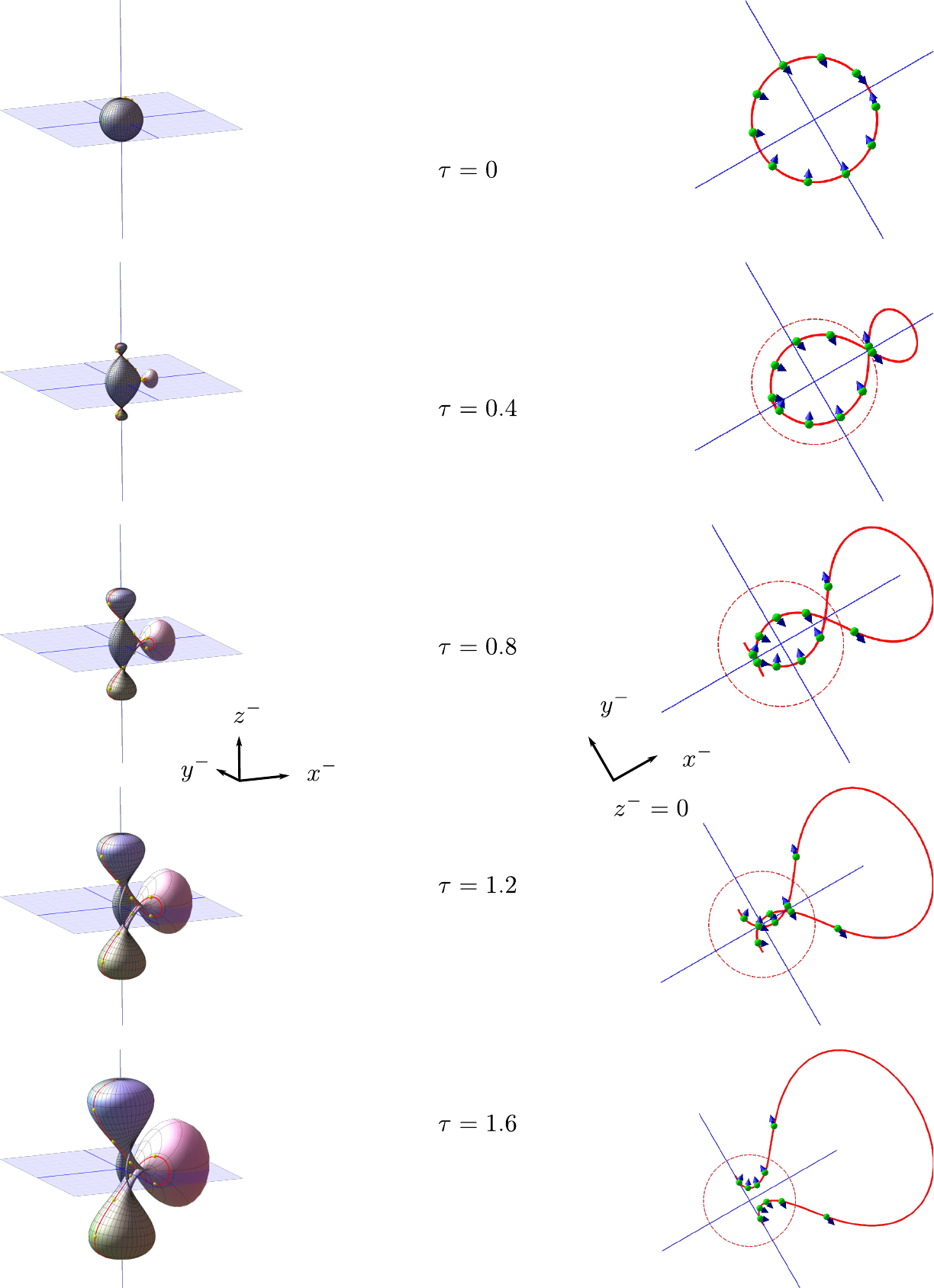}
    \caption{Deformation of an initially static swarm of particles that emerge instantly as a sphere of constant radius ${t_\ii^-}$. The impulse is generated by (\ref{KKS_fce_h_general}) with ${w_3=1=w_7}$ and string parameters ${\delta=\frac{1}{4}}$, ${\varepsilon=\frac{1}{8}}$. The asymmetry in the resulting motion of test particles is caused by the absence of one semi-infinite cosmic string along the negative $x$ axis and a boost induced by the second string creation.}\label{KKS_deformation}
\end{center}
\end{figure}

\begin{figure}[H]
\begin{center}
\includegraphics[keepaspectratio,scale=0.85]{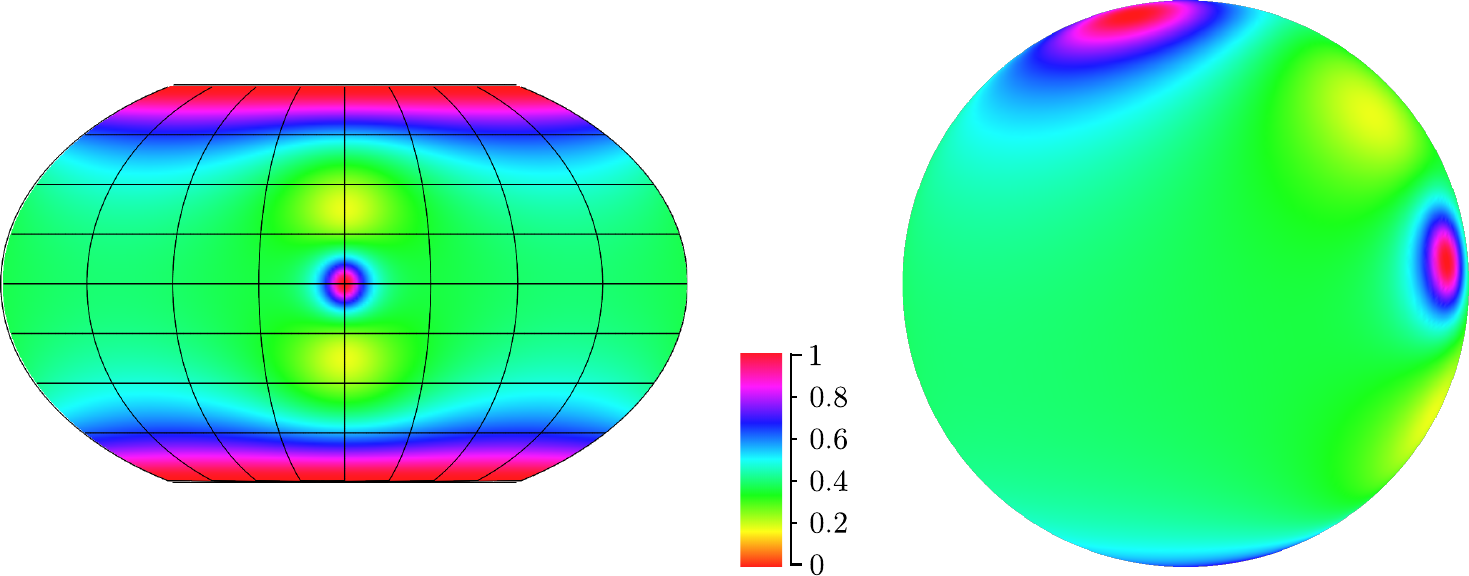}
\caption{In analogy with figure~\ref{fig:2st_velocities}, the distribution of velocity magnitudes (\ref{velocity_magnitude}) of initially static test particles, simultaneously escaping the impulsive surface, is plotted. Here, the impulse is generated by (\ref{KKS_fce_h_general}) with ${w_3=1=w_7}$ and ${\delta=\frac{1}{4}}$, ${\varepsilon=\frac{1}{8}}$. There are only three regions with a velocity magnitude approaching the speed of light corresponding to three moving string pieces, see the Kavrayskiy VII projection (left) and the spatial picture (right).}
\label{fig:2st_alt_velocities}
\end{center}
\end{figure}

Finally, we can visualize the kinematic effect of the impulse generated by (\ref{KKS_fce_h_general}) with ${w_3=1=w_7}$ on the motion of initially static test observers, see figure~\ref{KKS_deformation} for the evolution picture and figure~\ref{fig:2st_alt_velocities} for the velocity magnitude distribution. Due to the absence of a string piece along the negative $x$ axis and non-compensated inherent boost the resulting picture is asymmetric. Without an appropriate global view, one may be confused. In particular, restricting the analysis to just a quarter of the picture including two perpendicular string pieces in the positive $x$ and $z$ directions, and taking the deficit parameters ${\delta=\varepsilon}$ one would expect a symmetric picture in the presumed case of a pair of complete strings with four ends. However, we observe an induced motion of test particles that prefers the single-end direction.

\subsection{Two parallel strings}

\begin{figure}[H]
\begin{center}
\includegraphics[keepaspectratio,scale=0.95]{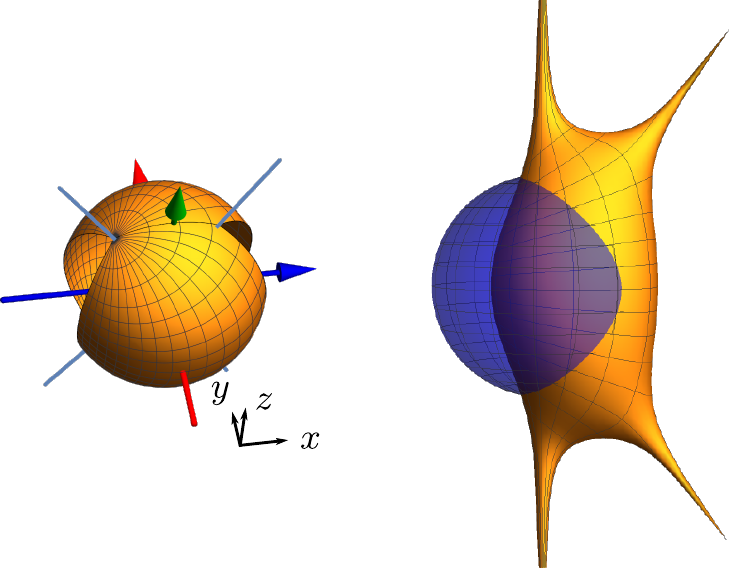}
    \caption{
    For the two parallel string configuration (\ref{2strings_par_fce_h_seq}) we plot the resulting unit Riemann sphere mapping in $\cal{M}^+$ (left). The string ends are visualised in light blue. Divergences of its image (right) in $\cal{M}^-$ indicated motion and mutual angles between these ends. In the right picture, we also plot the unit sphere in dark blue to represent the scaling.} \label{2strings_par_Riemann}
\end{center}
\end{figure}

As a last example let us briefly show ${h(Z)}$ describing \emph{two parallel strings}. By the definition, they cannot be made standing simultaneously. Therefore, the straightforward idea is to create the first (generically boosted) string, apply the rotation and another boost to induce its motion in the perpendicular direction, rotate back and create the second parallel string again with a generic boost. The relevant sequence in terms of elementary operations is
\begin{equation}    h(Z)=\rot{\pi/2,0,0}\boost{w_{10}}\rot{-3\pi/2,0,0}\cstring{\varepsilon}\rot{3\pi/2,0,0}\boost{w_5}\rot{3\pi/2,0,0}\boost{w_3}\cstring{\delta}Z \,. \label{2strings_par_fce_h_seq}
\end{equation}
with the string ends given by
\begin{equation}
    Z=\left\{0,\infty,\left(\frac{1}{w_3}\frac{1+w_5}{1-w_5}\right)^{\frac{1}{1-\delta}},\left(\frac{1}{w_3}\frac{1-w_5}{1+w_5}\right)^{\frac{1}{1-\delta}}\right\} \,.
\end{equation}
The mapping (\ref{2strings_par_fce_h_seq}) could be analysed in the same way as in two previous cases. Here, we only show the effect of such a mapping in terms of the Riemann sphere and deformation of the related ${t_\ii^+=\mbox{const}}$ spherical initial data, see figure~\ref{2strings_par_Riemann}.

\begin{figure}[H]
\begin{center}
\includegraphics[keepaspectratio,scale=0.95]{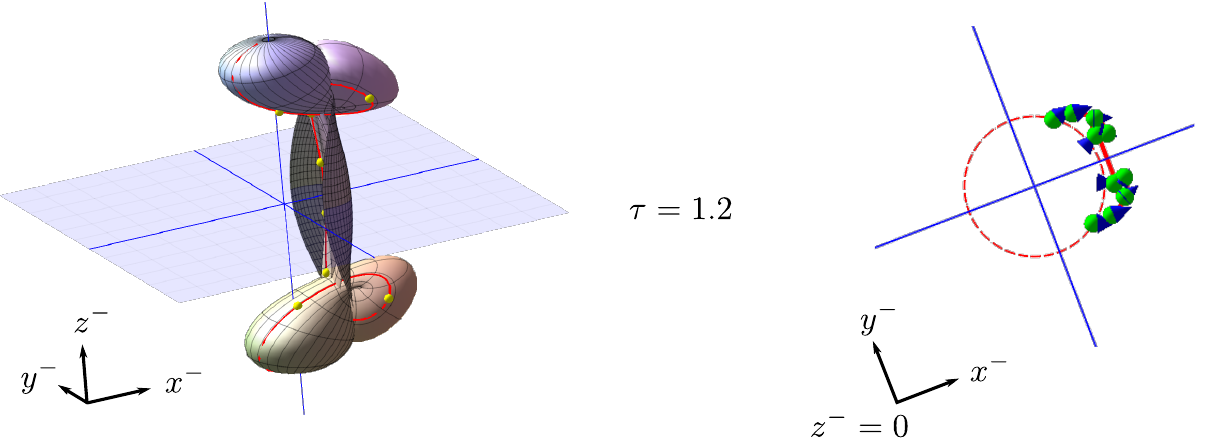}
    \caption{The initially static spherical distribution of test particles interacts with the expanding impulse generated by (\ref{2strings_par_fce_h_seq}) corresponding to a snapped pair of parallel cosmic strings. The only one typical evolution step of such a distribution along geodesics (\ref{flat_geodesics}) is depicted.} \label{2strings_par_motion}
\end{center}
\end{figure}

The physical interpretation of the above construction (\ref{2strings_par_fce_h_seq}) can be directly deduced from the resulting motion of initially static test particles, see figure~\ref{2strings_par_motion}. The deformation is again induced by the dragging of geodesics due to the motion of the string ends.

\section{Conclusions}

We studied geometries representing expanding impulsive gravitational waves. Our main aim was a geometric description and physical interpretation of the complex mapping $h(Z)$ entering the Penrose junction conditions (\ref{JuncCond}). In particular, situations related to the string-like nature of the wave source were elaborated. The $h(Z)$ properties and its refractive effects were connected with the motion of free test particles crossing the null wave surface. This was possible due to employing the recent rigorous results on the geodesic motion in expanding impulses. To clarify the role of elementary steps in $h(Z)$ construction, we analyzed specific initially static classes of geodesic congruences. Such an approach was introduced in the case of boosted one-string as the simplest possibility. Subsequently, it was followed by three situations with two strings, where we clarified and extended previous results. In particular, we studied a snapping pair cosmic strings, its degenerate subclass with only three string pieces generating the impulse, and finally, a case of two parallel strings. In general, the effects of expanding impulses generated by snapped cosmic strings acting onto geodesic motion can be described as a dragging of test particles by the string ends and induced motion in their directions. This analysis also showed non-trivial inherent boost-like effects within the construction of the two string scenarios and the way of its compensation.

\section*{Acknowledgements}
 DK acknowledges the support of the Czech Science Foundation, Grant 21-11268S. R\v{S} was supported by the Czech Science Foundation Grant No. GA\v{C}R 22-14791S.

\end{document}